\documentclass[12pt]{article}  
\usepackage{a4,latexsym,amsmath}
\usepackage[latin1]{inputenc}
\usepackage[T1]{fontenc}
\usepackage{amsfonts}
\usepackage{dsfont}
\usepackage[doublespacing]{setspace}
\usepackage[footnotesize]{caption}
\usepackage{graphicx}
\usepackage{natbib}
\usepackage{booktabs}
\usepackage{float}
\usepackage{wasysym}
\usepackage{ulem}
\usepackage[
  bookmarks=false,
  pdfpagelabels=false,
  hyperfootnotes=false,
  hyperindex=false,
  pageanchor=false,
  colorlinks,
]{hyperref}

\makeatletter 
\let\saved@hyper@linkurl\hyper@linkurl 
\let\saved@hyper@link@\hyper@link@
\AtBeginDocument{%
  \NoHyper
  \let\hyper@linkurl\saved@hyper@linkurl  
  \let\hyper@link@\saved@hyper@link@  
}
\makeatother

\usepackage{abstract}

\begin{document}
\normalem

\title{
\vspace{-50mm}
  \fbox{
  \begin{minipage}{\textwidth}%
  \begin{singlespace}
    \small{  
    \vspace{-5.5mm}
    This is the peer reviewed version of the following article: Hoshiyar, A., Kiers, H. A. L., \& Gertheiss, J. (2023). Penalized optimal scaling for ordinal variables with an application to international classification of functioning core sets.{ \it British Journal of Mathematical and Statistical Psychology},{ \it 00}, 1--19, which has been published in final form at
      \href{https://doi.org/10.1111/bmsp.12297}{https://doi.org/10.1111/bmsp.12297}.
      This manuscript version is made available under the
      \href{https://creativecommons.org/licenses/by-nc-nd/4.0}{CC-BY-NC-ND 4.0} license.}
      \end{singlespace}
  \end{minipage}
} \vspace{2mm}\\
  \textbf{Penalized Optimal Scaling for Ordinal Variables 
with an Application to International Classification of Functioning Core Sets
}
}

\maketitle

\begin{center}
\textbf{Abstract}
\end{center}
Ordinal data occur frequently in the social sciences. When applying principal component analysis (PCA), however, those data are often treated as numeric implying linear relationships between the variables at hand, or non-linear PCA is applied where the obtained quantifications are sometimes hard to interpret. Non-linear PCA for categorical data, also called optimal scoring/scaling, constructs new variables by assigning numerical values to categories such that the proportion of variance in those new variables that is explained by a predefined number of principal components is maximized. We propose a penalized version of non-linear PCA for ordinal variables that is a smoothed intermediate between standard PCA on category labels and non-linear PCA as used so far. The new approach is by no means limited to monotonic effects and offers both better interpretability of the non-linear transformation of the category labels as well as better performance on validation data than unpenalized non-linear PCA and/or standard linear PCA. In particular, an application of penalized optimal scaling to ordinal data as given with the International Classification of Functioning, Disability and Health (ICF) is provided.

\vspace{1cm}


\section{Introduction}\label{sectionIntro}

The objective of principal component analysis (PCA) is to reduce
the dimension of the data observed by finding a substantially smaller
set of linear combinations of the original variables --called \emph{principal components}-- that should
explain as much of the variability in the original data as possible. The principal components can, e.g., be obtained through an eigendecomposition of the (empirical) covariance/correlation matrix of the variables considered. If PCA is used as
an inferential tool, the data at hand should
follow (at least approximately) a normal distribution~\citep[cf.][Section 3.7]{Jolliffe:2002}. In practice, however, PCA is also widely employed as a descriptive/explorative tool without making distributional assumptions such as normality. Then the methodology is applicable to a wide variety of data. For instance, \citet{Labovitz:1970} proceeded by assigning numbers to rank order categories treating them as interval scaled and demonstrated via simulations that the resulting errors can be negligible under some circumstances. Nonetheless, since the idea of (Pearson) correlation has been developed for data of metric nature, caution is warranted when analyzing categorical data. \citet{Kolenikov:2004} discuss various methods dealing with categorical data in the context of principal component analysis in an extensive simulation study. Their proposed technique of using polychoric correlations, might indeed be preferable if the statistical properties of the PCA model are of primary interest, but the method is computationally quite intensive. \citet{Korhonen:1998} suggest an approach called `ordinal principal component analysis' with rank correlations to be maximized between the original variables and the ordinal principal component. Again, the suggestion suffers from both high constructional and computational effort for larger data sets. Another procedure is the more established concept of non-linear PCA \citep{Gifi:1990, Mori:2016}. The basic idea of non-linear PCA is to build linear combinations of non-linear transformations of the original variables in order to further increase the variance that is explained by the first principal components.

Here we also consider ordinal variables, that is, categorical variables with levels that can be reasonably ordered. Such variables are often found in the social and behavioral sciences. Though many practitioners simply treat category labels as numeric values and apply standard PCA (along the lines of \citet{Labovitz:1970}), this way of analysis may still be considered questionable, since ordinal variables do not have metric scale level. Preferably, variables should be treated at their appropriate measurement level to avoid overestimating the information contained. Furthermore, just as for regression analysis, situations are imaginable, in which it is reasonable to assume the relationships between the variables to be non-linear. Non-linear PCA was designed to take those considerations into account. Non-linear PCA for categorical data constructs new variables by assigning numerical values to categories such that the proportion of variance in those new variables that is explained by the first, lets say $m$, principal components is maximized. This process is called `optimal quantification', `optimal scaling' or `optimal scoring'; cf.~\citet{LinMeuKooGro:2007} and references therein. However, while the found transformations --the `quantifications'-- maximize the explained variance on the data at hand, the `training data', it is by no means clear that they will also work well on new data, sometimes called `validation data' or `test data'. In fact, by simply maximizing the explained variance on the training data, the found transformations may rather account for random fluctuations in the data than for substantial non-linearity. In other words, the estimated scores tend to result in `overfitting' the (training) data, which worsens the performance and generalization to new data. In addition, the obtained quantifications are sometimes erratic and thus hard to interpret.

To attack those problems, we propose a penalized version of non-linear PCA for ordinal variables. In other settings, particularly regression analysis, penalty methods have already proven to provide a valuable means to both reduce overfitting and increase interpretability of the results, while at the same time circumventing the restrictions implied by strictly linear modeling. To the best of our knowledge, however, penalization has never been proposed, discussed, nor used in the context of optimal scaling/non-linear PCA as done here. When using penalization in the framework of PCA or explanatory factor analysis, literature has so far typically been on shrinking loadings obtained via linear PCA toward zero to avoid the drawbacks of the usual hard-threshold approach; see, e.g., \citet{Zou:2006}, \citet{WittenEtal:2009}, \citet{Jin:2018}. Our framework, motivation and approach is very different. We are targeting the quantifications, i.e., the scores assigned to the levels of ordinal variables within optimal scaling, not the factor loadings (within linear PCA). Our approach hence provides novelty by introducing penalization to a popular and well-established framework --optimal scaling-- in a way it has not been considered before. So far, `penalized/regularized optimal scaling/scoring' as found in the literature~\citep{Hastie:1994,Hastie:1995,Meulman:2019} refers to a supervised learning problem (discriminant analysis/regression, not PCA) and, more importantly, imposes a penalty on the $\beta$-coefficients of the (scaled) predictors, not the scoring functions.

The PCA-approach for ordinal variables that we propose here is a smoothed intermediate between standard PCA on category numbers and non-linear PCA (without smoothing) as typically used so far. Specifically, we extend the idea of quadratic second-order penalties presented in \citet{GerOeh:2011} and \citet{Gertheiss:2022} in the context of regression with ordinal predictors to the framework of optimal scaling. The motivation behind using second-order penalties is that in the extreme case of maximum penalty the result is common (linear) PCA on category numbers. On the other hand, with vanishing penalty, the standard optimal scaling approach as described above is obtained. Between those two extremes, however, the estimated quantifications are still non-linear but smoother than the scores resulting from purely/standard non-linear PCA/optimal scaling. The specific strength of the penalty, acting as a tuning/smoothing parameter, and hence the concrete form of `intermediate' needs to be chosen in a data-driven manner, along with the number of components to extract. Methods for doing so and potential pitfalls will be discussed here as well. In summary, the main motivation and goals of the new approach are to offer both better interpretability of the non-linear transformations (i.e., the scoring/quantification rules) within optimal scaling as well as better performance on validation data than unpenalized/usual optimal scaling for ordinal data and/or standard linear PCA. Also, our implementation provides the option of both non-monotone effects and incorporating constraints enforcing monotonicity with respect to the scores assigned, as the latter assumption is reasonable in some practical applications. Based on a related idea of penalizing large differences between adjacent categories, \citet{Burkner:2020} proceeded from a Bayesian point of view to model monotonic effects of ordered independent variables in regression analysis. The only existing method for smoothing within optimal scaling with categorical data we are aware of is available as an option in IBM SPSS Statistics \citep{CATPCA}. Here, scores can be fitted as spline functions, with the number of (interior) knots being chosen by the user. On the one hand, however, smoothing by using a small and manually chosen number of knots (the default in SPSS is `2') limits the type of functions that can be fitted and may be challenging for the (inexperienced) user (without further guidance on how to choose the number and placing of knots). That is why, state-of-the-art methods for regression splines rather use a large number of basis functions/knots in combination with a smoothing penalty, and a smoothing/penalty parameter that is determined by using a specific, data-dependent criterion such as (generalized) cross-validation; see, e.g., \citet{EilMar:1996}, \citet{Wood:2008, Wood:2017}. On the other hand, splines are defined on a sub-interval of $\mathds{R}$, whereas ordinal variables --the focus of our studies-- can only take some discrete values. In other words, a spline function may be seen as unnecessarily complex for scaling ordinal/discrete data. Consequently, a smoothing penalty that targets specifically the discrete levels and assigned scores of an ordinal variable appears to be a more sensible approach for smoothed optimal scaling with ordinal data. This is exactly what will be presented in this article.  

A representative example for ordinal variables is the so-called International Classification of Functioning, Disability and Health (ICF; \citealp{WHO:2001}) core set for chronic widespread pain (CWP). Besides observed levels of the 67 ICF variables, the data set at hand~\citep{Cieza:2004, GerHogObeTut:2011} also provides a physical health component summary measure, originally constructed from the SF-36 questionnaire by standard, linear PCA \citep{Mchorney:1993}. The proposed penalized, non-linear PCA seems promising to allow a more intuitive way to derive the overall health condition in an unsupervised fashion directly from the ICF data, which constitutes another substantial contribution of this article. 

The remainder of the paper is organized as follows. Section~\ref*{Sec2} provides a detailed explanation of optimal scaling and the proposed penalized fitting algorithm. Some illustrative simulation studies in Section~\ref*{Sec3} examine the performance of our new scoring procedure under the presence and even absence of monotonicity. In Section~\ref*{Sec4}, we consider the ICF case study dealing with measures of CWP; also, further results are presented on another publicly available data set~\citep{JouEtal:1988}. In Section \ref*{Sec5} we conclude with a discussion.  
All computations were done using the statistical program \texttt{R} \citep{R}.
To ensure reproducibility, the algorithm together with evaluation on publicly available data as given above is made accessible through open source \texttt{R} add-on package \texttt{ordPens}~\citep{ordPens:2021}.

\section{Penalized Optimal Scaling}\label{Sec2}

Before introducing our approach for penalized non-linear PCA with ordinal variables, we will first shortly review standard linear PCA and optimal scaling for categorical variables. For proofs and mathematical derivations on principal components, we refer to \citet{Jolliffe:2002}, as a more detailed introduction to linear PCA would lie beyond the scope of the present work. For background on non-linear PCA and its alternating algorithm, see \citet{Mori:2013}. If particularly interested in interpreting non-linear PCA models, we refer to \citet{LinMeuKooGro:2007}.

\subsection{Linear PCA and Optimal Scaling}\label{SubSec21}
As usual in multivariate statistics, we assume that the observed
data matrix $X$ has entries $(X)_{ij} = x_{ij}$ denoting the value
of the $j$th variable observed at the $i$th subject, $j=1,\ldots,p$,
$i=1,\ldots,n$. Furthermore, we assume that $p < n$ and variables to be centered
and scaled to have unit variance (with the usual divisor of $n-1$). Then, the empirical covariance/correlation matrix is
given by $R = (n-1)^{-1}X^\top X$. Informally speaking, our goal is to extract the important information from the correlated data, i.e., to achieve dimension reduction with as little loss of information as possible.

The idea of principal component analysis is to find linear
combinations $y_r = Xa_r$, with the variation in the original data
that is explained by the first $m$ principal components being as large as possible; $a_1,\ldots,a_m$ are the corresponding vectors of `loadings'. Specifically, this means that the proportion of \emph{`variance accounted for'}
\begin{equation}\label{def:VAF}
\text{VAF} = \frac{\sum_{r=1}^m y_r^\top
	y_r}{\sum_{j=1}^p x_j^\top x_j}
\end{equation}		
has to be maximized, with
$x_j=(x_{1j},\ldots,x_{nj})^\top$, $m < p$, and subject to some constraints. First, the principal components need to be uncorrelated, which can be interpreted such that no component should contain information that is already provided by another component. Second, as the numerator
depends not only on the directions but also on the lengths of the vectors $a_r$, those $a_r$ are restricted to have unit length, i.e., $a_r^\top
a_r = 1$ for all $r$. Please note, technically, when defining quantity/VAF (\ref*{def:VAF}) to be maximized, the denominator $\sum_{j=1}^p x_j^\top x_j$ would not be needed, because it does not depend on $a_1,\ldots, a_m$ but gives the overall variation in the original data (which equals $(n-1)p$ if the data is standardized). However, it ensures that VAF is between zero and one, and can hence be interpreted as a `proportion'. This (constrained) optimization problem is solved by the first $m$ (orthonormal) eigenvectors $a_1,\ldots,a_m$ of correlation matrix $R$ which correspond to the $m$ largest eigenvalues $\nu_1 \geq \ldots \geq \nu_m$.
As an alternative to the eigendecomposition, the same vectors can be found through a singular value decomposition (SVD) of $X$. Here, we follow the latter representation in accordance with \texttt{R} function \texttt{stats::prcomp}, which is also part of our algorithm's implementation. 
 SVD means that $X$ is decomposed in terms of
 $X = U D V ^\top$, 
with $U$ and $V$ being $(n \times p)$ and $(p \times p)$ matrices, respectively, where the columns of $V$ correspond to eigenvectors $a_r$. $D$ is a diagonal $(p \times p)$ matrix and contains the square roots of eigenvalues $\nu_r$, sorted decreasingly.

With ordinal variables, matrix $X$ (before standardizing) contains only integers $1,2,\ldots$,
with entry $x_{ij}$ indicating the level of the $j$th variable that is observed at the $i$th
subject. As numbers $1,2,\ldots$ are just class labels, linearity in these labels as assumed
by usual PCA appears to be very restrictive and not necessarily the right choice for categorical data. Therefore, non-linear PCA constructs new variables by assigning numerical values to categories in terms
of $\phi_{ij} = \varphi_j(x_{ij})$, with scaling functions $\varphi_j : \mathds{N} \rightarrow \mathds{R}$, $j=1,\ldots,p$.
Then, standard PCA as described above is applied to the recoded variables. As before with data matrix $X$, recoded variables are standardized to have mean zero and variance one. To find appropriate, or `optimal' functions $\varphi_j$ to use for recoding, the proportion of variance in the transformed variables that is explained by the first $m$ principal components is maximized, with $m$ being fixed at a certain value. For formalizing the optimal scaling problem, we take advantage of a well-known property of the loadings 
$\hat{a}_1,\ldots,\hat{a}_m$, $\hat{a}_r=(\hat{a}_{1r},\ldots,\hat{a}_{pr})^\top$, from standard/linear PCA, and corresponding vectors of component scores $\hat{y}_1,\ldots,\hat{y}_m$,
$\hat{y}_r=(\hat{y}_{1r},\ldots,\hat{y}_{nr})^\top$, compare~\citet{Young:1978}.
Namely, that they solve the problem of minimizing $L(Y, A)$
 in terms of $\{\hat{Y},\hat{A}\} = \arg\min_{Y,A} L(Y,A)$, with
\begin{equation}
L(Y,A) = \sum_{j=1}^p\sum_{i=1}^n \left(x_{ij} - \sum_{r=1}^m y_{ir}a_{jr}\right)^2 = \| X - YA^\top \|_F^2,
\end{equation}
$\|\cdot \|_F$ denoting the Frobenius norm,
$Y$ being the ($n\times m$) score matrix with entries $(Y)_{ir} = y_{ir}$ and $A$ being ($p \times m$) with entries $(A)_{jr} = a_{jr}$, $r = 1,\ldots,m$,
$j=1,\ldots,p$, $i=1,\ldots,n$. That means, the vectors $a_1,\ldots,a_m$ are the columns of $A$, the matrix of loadings.
For non-linear PCA, criterion
\begin{equation}
L(\Phi,Y,A) = \sum_{j=1}^p\sum_{i=1}^n \left(\phi_{ij} - \sum_{r=1}^m y_{ir}a_{jr}\right)^2
\end{equation}
is minimized as a function of matrices $A$, $Y$ and $(n \times p)$ matrix $\Phi$, with $(\Phi)_{ij} = \phi_{ij} = \varphi_j(x_{ij})$;
see \citet{LinMeuKooGro:2007}. Now, $\hat{A}$ and $\hat{Y}$ correspond to the matrix of loadings and
respective PC scores when using the transformed variables. Scaling function $\varphi_j$ can also
be determined by the vector $\theta_j = (\theta_{j1},\ldots,\theta_{jk_j})^\top$ where $\theta_{jl}$
is the value that is assigned to category $l$ of the $j$th (categorical) variable, and $k_j$ denotes
the highest level of variable $j$. Then,  $\phi_{ij} = \varphi_j(x_{ij}) = z_{ij}^\top\theta_j$, with $z_{ij} = (z_{ij1},\ldots,z_{ijk_j})^\top$ being a
design vector of dimension $k_j$ with entry $z_{ijl} = 1$ if at subject $i$ variable $j$ has
value $l$, and zero otherwise.

For fixed quantifications $\Phi$, $L(\Phi, Y, A)$ is minimized by the usual PCA solution on
data matrix $\Phi$ (note, we just replaced $X$ by $\Phi$). For fixed $Y$ and $A$, minimization
of $L(\Phi,Y,A)$ becomes a ``regression problem'' in terms of $\hat{\theta} = \arg\min_{\theta} Q(\theta,Y,A)$, with
\begin{equation}\label{eqLPhi}
Q(\theta,Y,A) = \sum_{j=1}^p\sum_{i=1}^n (u_{ij} - z_{ij}^\top\theta_j)^2,
\end{equation} 
coefficient vector $\theta = (\theta_1^\top,\ldots,\theta_p^\top)^\top$ and pseudo response $u_{ij} = \sum_{r=1}^m y_{ir}a_{jr}$. The complete indicator matrix is given by
$Z = (Z_1|\ldots|Z_p)$ and
\begin{equation}\label{eqZj}
Z_j = (z_{ijl}) =
\begin{pmatrix}
z_{1j1} & \cdots & z_{1jk_j}\\
 \vdots &  \ddots & \vdots \\
 z_{nj1} & \cdots & z_{njk_j} \\
\end{pmatrix}.
\end{equation}
Then, the estimated quantifications are obtained as
$$\hat\Phi = \underbrace{(Z_1|\ldots|Z_p)}_{Z}
\begin{pmatrix}
\hat\theta_1 & & \\ 
& \ddots & \\
& & \hat\theta_p \\
\end{pmatrix}. 
$$
That means, the $j$th column of $\hat{\Phi}$, containing the quatifications of the $j$th (ordinal) variable, is given through $Z_j\hat\theta_j$. If it is reasonable to assume that changes between adjacent categories arise consistently in up- or downward fashion, monotonicity constraints are to be imposed. In the case of monotonically increasing quantifications, for instance, this can be achieved by introducing the $((k_j-1)\times k_j)$ difference matrix of first order $D_1$ such that the differences
$$
D_1 \theta_j =
\begin{pmatrix}
-1 & 1 & & & \\
& & \ddots & \ddots & \\
& & & -1 & 1 \\
\end{pmatrix}
\begin{pmatrix}
\theta_{j1} \\
\vdots \\
\theta_{jk_j} \\
\end{pmatrix}
=
\begin{pmatrix}
\theta_{j2} - \theta_{j1} \\
\vdots \\
\theta_{jk_j} - \theta_{jk_{j}- 1} \\
\end{pmatrix}
\geq
\begin{pmatrix}
0 \\
\vdots \\
0 \\
\end{pmatrix}
$$ are enforced to be non-negative. For finding the final solution, $\Phi$ and $\{Y,A\}$ are alternately fixed at their
current value, and it is cycled through the two optimization steps until convergence.
For further details on the alternating least squares (ALS) algorithm, see \citet{Young:1978}.
In order to incorporate constraints, such as monotonic quantifications, we make use of the dual routine of Goldfarb and Idnani~(\citeyear{Goldfarb:1982}, \citeyear{Goldfarb:1983}) on solving quadratic functions under constraints, as implemented in \texttt{R} add-on package \texttt{quadprog}~\citep{quadprog}.
 
For selecting the number of principal components, and thus an appropriate value for $m$, the same methods as for usual PCA can be used; compare, e.g., \citet{LinMeuKooGro:2007, Manisera:2010, Linting:2012}. Note, once the scoring rules have been fit (using some $m$), optimal scaling is nothing else than usual, linear PCA on the scaled data (i.e., after applying the estimated scoring/scaling functions $\hat\varphi_j$ to the data). A popular way to choose the number of components is the so-called scree-criterion/plot \citep{Catell:1966}, where the PCs are plotted against their eigenvalues, i.e., variances, in order to detect the first break (`elbow'). Since, however, different values of $m$ may lead to different scaling functions, the PCA solution on the scaled data, and hence the eigenvalues, may look different for different $m$ (the nonlinear PCA solutions are not nested). Therefore the scaling functions should be fit for various $m$-values and the corresponding scree-plots should be compared to have the full picture. It has been our experience though that very often various $m$-values eventually lead to the same number of components as resulting from the scree-criterion. Then, of course, this specific number should be used for $m$; or in other words \citep{LinMeuKooGro:2007}: ``If a different number of components than $m$ is chosen, the nonlinear PCA should be rerun with the chosen number of components because the components are not nested''. As an alternative to the scree plot, we could also use the popular criterion/strategy to choose the number of extracted components, resp.~$m$ value, as the minimum value such that (at least) a pre-defined proportion of variance is explained. It should be noted though, that this proportion is often set to be around 90\%, which typically leads to a very large number of compenents even for moderate $p$ if correlations between the (scaled) variables are not extremely high.
  
\subsection{Penalized Fitting}\label{Subsec22}
When $Q(\theta,Y,A)$ at (\ref*{eqLPhi}) is minimized as a function of $\theta_1,\ldots,\theta_p$, only the nominal scale level of the variables is used (unless monotonicity constraints are set). In regression problems, it has been proposed to use special penalties to incorporate the covariates' ordinal scale level; see, e.g., Tutz and Gertheiss~(\citeyear{Tutz:2014}, \citeyear{Tutz:2016}) for an overview. Similar approaches can be used here. In particular, penalizing non-linearity in the coefficients as done in \citet{GerOeh:2011} seems promising. Due to the fact that categories considered here are ordered, changes between adjacent levels can be assumed to take place rather smoothly. In order to avoid abrupt jumps between levels, we hence smooth out the coefficient vector. So the idea is not to minimize $Q(\theta,Y,A)$ at (\ref*{eqLPhi}) as a function of $\theta = (\theta_1^\top,\ldots,\theta_p^\top)^\top$, but its penalized version
\begin{equation}\label{eqLpPhi}
Q_p(\theta,Y,A) = \sum_{j=1}^p\sum_{i=1}^n (u_{ij} - z_{ij}^\top\theta_j)^2 +  \sum_{j=1}^p \lambda_j J_j(\theta_j),
\end{equation}
with $\lambda_j = \lambda(k_j-1)$. For the penalty terms $J_j(\theta_j)$, we choose
\begin{equation}\label{defJ}
J_j(\theta_j) = \sum_{l=2}^{k_j-1} ((\theta_{j,l+1} - \theta_{jl}) - (\theta_{jl} - \theta_{j,l-1}))^2 =
\sum_{l=2}^{k_j-1} (\theta_{j,l+1} - 2\theta_{jl} + \theta_{j,l-1})^2.
\end{equation}
To be more concise, equation~(\ref*{defJ}) takes the general form $J_j(\theta_j) = \theta_j ^\top \Omega_j \theta_j$
with $(k_j \times k_j)$ penalty matrix $\Omega_j = D_2 ^\top D_2 $ and $((k_j-2)\times k_j)$ second-order difference matrix
$$
D_2 =
\begin{pmatrix}
1 & -2 & 1 & & & \\
& 1 & -2 & 1 & & \\
& & \ddots & \ddots & \ddots & \\
& & & 1 & -2 & 1 \\
\end{pmatrix}.
$$
By using this quadratic, second-order penalty, we penalize non-linearity in the $\theta$-coefficients that belong to the same variable.  

The strength of penalization is controlled by parameter $\lambda$. With $\lambda = 0$, optimal scaling for categorical variables as described above is obtained; with $\lambda \rightarrow \infty$, coefficients are forced to be linear, which is equivalent to usual PCA using (standardized) class labels $1,2,\ldots,k_j$ for variable $j$. With $0 < \lambda < \infty$, coefficients are non-linear but smoother than with unpenalized non-linear PCA/optimal scaling, which makes good sense for ordinal variables, as wiggly coefficient vectors $\theta_j$ are hard to interpret. How to choose an appropriate value for $\lambda$, will be discussed in Subsection~\ref*{lamopt} below. 

\subsection{The Penalized  ALS Algorithm}
For initializing the algorithm, $X$ itself serves as initial data matrix/matrix of quantifications denoted by $\hat\Phi^{(0)}$ (and is standardized columnwise if this has not been done already). Further let $\hat\phi_j^{(t)}$ denote the $j$th column of $\hat\Phi$ in iteration $t$ of the algorithm, i.e., $\hat\Phi^{(t)}$. Then, we iteratively cycle through an PCA step where PCA is carried out for fixed quantifications/scaling functions, and a quantification step where the (smoothed) scaling functions $\varphi_j$ are estimated for fixed loadings $A$ and scores $Y$ as motivated above. Specifically, we have:
\begin{itemize}
\item \textbf{PCA step}: Calculate $\hat{A}^{(t+1)}$ and $\hat{Y}^{(t+1)}$ through SVD of the matrix of current quantifications $\hat\Phi^{(t)}$ and update the matrix of pseudo responses
                         $U^{(t+1)} = \hat{Y}^{(t+1)} \hat{A}^{(t+1)\top}$.
\item \textbf{Quantification step}: Columnwise, calculate estimates/coefficient vectors $\hat\theta_j^{(t+1)}$ by minimizing (\ref*{eqLpPhi}) using the pseudo responses/entries $u_{ij}$ of $U^{(t+1)}$ obtained from the PCA step above, design/indicator matrices $Z_j$ as defined in~(\ref{eqZj}), and subject to constraints (i) standardization/unit variance w.r.t.~$\hat\phi_j^{(t+1)} = Z_j \hat\theta_j^{(t+1)}$,  and (ii) monotonicity (if so) w.r.t.~$\hat\theta_j^{(t+1)}$. (i) is realized by use of the estimate from the iteration before, more precisely, optimization is done subject to constraint $(\tilde\phi_j^{(t)})^\top Z_j\theta_j^{(t+1)} = 1$, where $\tilde\phi_j^{(t)}$ is the (standardized) estimate from iteration $t$. (ii) is done as described above by use of matrix $D_1$ of appropriate dimension. Update the estimated quantifications of the $j$th variable $ \hat\phi_j^{(t+1)} = Z_j \hat\theta_j^{(t+1)}$. 
\end{itemize}
Once a user-specified convergence criterion has been reached, apply a final (standard) PCA to the quantified data matrix. The convergence criterion we use is
$$ 
\frac{1}{np}\sum_{i=1}^n\sum_{j=1}^p ( \hat\phi_{ij}^{(t)} - \hat\phi_{ij}^{(t+1)} )^2 < \epsilon,
$$
where $(\hat\Phi^{(t)})_{ij} = \hat\phi_{ij}^{(t)}$, and $\epsilon$ is a small, positive constant such as $\epsilon = 10^{-7}$. At this point we would like to highlight the importance of standardizing the data matrix/matrix of quantifications before applying any PCA. Otherwise, the optimal scaling algorithm would just assign arbitrarily large values to the levels of an arbitrary set of $m$ variables, while shrinking the remaining variables towards zero, because then, trivially, 100\% variance would be explained by $m$ principal components. This `solution' of the optimization problem supposed to be solved through optimal scaling, however, would obviously be neither reasonable nor useful.

\subsection{Selection of the Smoothing Parameter}\label{lamopt}
In the algorithm above, the penalty parameter $\lambda$ was fixed at some specific value. The choice of an optimal value for $\lambda$, however, should be made using the data at hand. Common strategies in penalized regression involve information criteria or cross-validation techniques. If PCA is used as an exploratory tool without distributional assumptions as done here, $K$-fold cross-validation can be used, since it does not require a likelihood. The general procedure is that the data is randomly split into $K$ folds/subsets $k=1,\ldots,K$ of similar size. Given the $k$th subset is used as the so-called validation set, unknown parameters are fitted on the remaining $K-1$ parts of the data (the so-called training set). Using those parameters, an appropriate measure of performance $M_k$ is calculated on the $k$th part of the data. Specifically, with penalized non-linear PCA as proposed here, the parameters of interest are the scaling functions. So we use the quantifications as estimated on the training set to scale the validation data (the $k$th subset). Then, we run a (standard) PCA on the scaled validation data and use the proportion of variance that is explained by the first $m$ principal components as the measure of performance $M_k$ of the fitted scaling function. Recall, the basic idea behind optimal scaling is to maximize the variance that is accounted for/VAF~(\ref*{def:VAF}) by the first $m$ principal components. For instance, if the scaling function rather fits noise in the training data than results from substantial non-linearity, it will perform worse on independent validation data than a simpler, rather linear function. Since the quantifications in our case depend on the value of the penalty parameter $\lambda$, $M_k$ also depends on $\lambda$, and is hence written as $M_k(\lambda)$. Then, cycle through all $k=1,\ldots,K$ partitions and calculate 
$$
\text{CV}(\lambda)= \frac{1}{K} \sum_{k=1}^K M_k(\lambda). 
$$ 
Now, over a fine grid $G$ of sensible values $\lambda \in G$, the optimal $\lambda$ can be determined by maximizing the cross-validated VAF.

Cross-validation, however, can be time-consuming, since parameters need to be fitted repeatedly. Interestingly, in many practical applications of penalized non-linear PCA, also a simple graphical tool can be used to find an appropriate value for $\lambda$. 
Instead of cross-validated VAF, for each candidate value of $\lambda$ the parameters are only fitted once, using the entire data set as training data. Then, the VAF on the training data is drawn as a function of $\lambda$. By definition of penalized non-linear PCA, this function, VAF$(\lambda)$, is monotonically decreasing (as with a smaller $\lambda$ more emphasis is put on the data). However, it is often found that this function is almost constant for $\lambda$ below some $\lambda_0$ (and for $\lambda$ above some $\lambda_1$). Consequently, $\lambda_0$ may be used as a good compromise between fit to the data and interpretation. In other words, with $\lambda < \lambda_0$ we run into `over-fitting', where more pronounced non-linearity (obtained for smaller $\lambda$) only leads to marginal improvement with respect to VAF. With $\lambda > \lambda_0$, on the other hand, we observe a substantial drop in VAF. Examples where this behavior of VAF$(\lambda)$ is observed, are found in Section~\ref*{Sec4}. 

As an alternative to the purely graphical approach, we can, e.g., choose $\lambda_0$ as the largest $\lambda$  
still fulfilling the condition
$$
\text{VAF}(0) - \text{VAF}(\lambda) = \frac{1}{p} \sum_{r=1}^m \nu_r(0) - \frac{1}{p} \sum_{r=1}^m \nu_r(\lambda)  \leq \delta, 
$$
where $\nu_r(\lambda)$ denotes the $r$th (largest) eigenvalue obtained in the final PCA with penalty parameter $\lambda$, and $\delta$ being a small (preselected) constant, e.g., $1 \permil$; VAF$(\lambda)$ denotes the corresponding VAF on the training data.

Once an appropriate penalty parameter has been found for a specific $m$ value, we can run penalized optimal scaling with this specific $\lambda$ and proceed to choose the number of principal components as described at the end of Subsection~\ref{SubSec21} above. If, e.g., scree plots are to be drawn for different values of $m$, we should of course select corresponding/optimal $m$-specific $\lambda$ values first. An illustration and discussion of this approach for the ICF data is found in Section~\ref{Sec4} below.

\section{Numerical Experiments}\label{Sec3}
Before applying penalized non-linear PCA as proposed to real world data, we carry out some illustrative numerical experiments to confirm that the method is able to identify the underlying structure used for data generation. This allows us to gain some insight into statistical properties such as the so-called bias-variance trade-off, which is commonly seen for smoothing techniques (compare, e.g., \citet{Fahrmeir:2001}). To be concrete, a larger value of the penalty parameter is typically associated with a larger bias and comparatively low variance, and vice versa. A related aspect of interest is whether and to what extent the procedure is qualified to detect relationships beyond linearity. Note, however, that the term `bias' is rather used informally/qualitatively speaking here, since there is no `true' scoring rule the obtained quantifications could be compared to in a strict sense by taking differences (mainly due to the method of creating ordinal observations from continuous data by thresholding, see below).

\subsection{Experimental Design}
For illustration and evaluation of our method, we conducted this simulation with varying design levels of the smoothing parameter, sample size and the standard deviation of noise overlaying the data obtained as linear combinations of some latent factors. For $\lambda$ we opt for different values according to no penalization (purely non-linear PCA), small, medium and large penalization, with the latter tending towards standard, linear PCA. We chose a rather large sample size of $n=500$, which is comparable to our empirical example on the ICF data, along with a quite small value of $n=100$ to illustrate the method's potential limitations. In the latter case, we ran the algorithm with $\lambda = 0, 0.1, 2, 5$. 
Since the effect of the penalty tends to vanish off as the sample size 
exceeds a certain value, the smoothing parameters are adjusted to $\lambda = 0, 0.1, 5, 10$ in the case of $n=500$. The number of variables ($p = 20$) and the number of principal components/latent factors ($m = 5$) are kept fixed. While varying one parameter ($\lambda$ or $n$), we keep the other constant such that we observe a total of $8$ scenarios. 
For reasons of simplicity, we chose a design assuming a five-point scale for each variable. For the eigenvalues $\nu_r$, $r = 1,\ldots,5$, which correspond to the variances of the underlying factors, we assume the descending sequence $6, 5, 4, 3, 2$. The (true) loadings matrix $A$ is chosen such that each component loads on a distinct set of variables; more precisely

\[
\underbrace{A}_{20 \times 5} = \begin{pmatrix}
a_{(1)} &  &  \\
 & \ddots & \\
 &  & a_{(5)} \\
\end{pmatrix},
\]

with $a_{(r)}$ being a normalized $((7-r)\times 1)$ vector with entries $1/\sqrt{7-r}$. By doing so, we achieve a sparse matrix, having a positive loading on one factor only for each variable, which is ideally wished for factor interpretation in practice.  
Matrix $A$ doesn't change over scenarios. For generating the five latent factors, we draw i.i.d.~normal data, with zero means and variances corresponding to the eigenvalues given above. The resulting $(n \times 5)$ (latent) data matrix $\tilde{Y}$ is multiplied by $A^\top$, and overlaid with i.i.d.~gaussian error with variance $\tau^2$. The resulting matrix, say  $\tilde{X} = \tilde{Y} A^\top + E$, is then discretized  by applying cut-points (similarly to a threshold mechanism known from cumulative ordinal regression models). Those cut-points are chosen such that five different types of quantifications are obtained, involving monotonic as well as non-monotonic effects: V-shaped (variables 1--6), S-shaped (variables 7--11), linear (variables 12--15), square root (variables 16--18) and quadratic (variables 19--20). For simplicity, the same transformation is applied to each variable within the same PC (compare matrix $A$ above), leading to a construction where the V-shaped variables load on PC 1, S-shaped on PC 2, and so on.  
In case of a monotone (i.e., S-shaped, linear, square root and quadratic) scoring rule, cut-points are simply obtained by applying corresponding transformations on some equidistant grid-points.   
For the non-monotonic quantifications, we use two cut-points $\zeta_1$ and $\zeta_2$ only, with latent/continuous observations falling into $[\zeta_1,\zeta_2]$ being randomly assigned to level 2 or 4, and observations above $\zeta_2$ being denoted as level 1 or 5 (also chosen at random). The remaining data points (i.e., below $\zeta_1$) are interpreted as level 3. The motivation/interpretation behind this is that the latent factor merely determines whether extreme categories are observed or not. 
The entire process of data generation, transformation/discretization, and penalized optimal scaling was repeated 500 times. Result are discussed below.

\subsection{Results} 
To check whether the method presented in Section~\ref*{Sec2} is qualified to recover the underlying transformations of the data, we will examine the estimated quantifications $\hat{\theta}$ for different values of penalty parameter $\lambda$, given a certain sample size $n$ and error variance $\tau^2$.
   
\begin{figure}[!ht]\centering
\includegraphics[width=\linewidth]{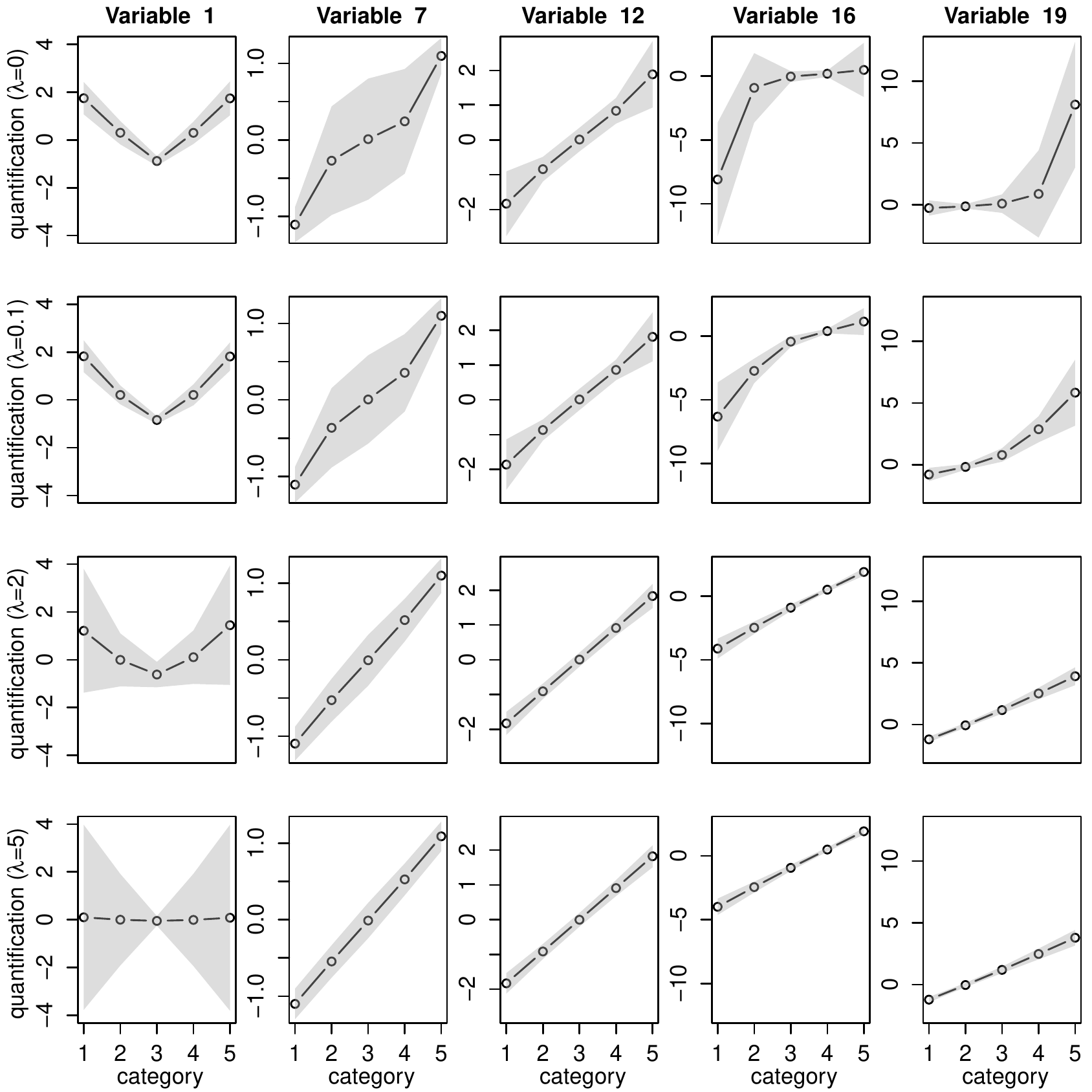}  
\caption{\label{Figure_1} Estimated quantifications (means, black lines) and pointwise uncertainty bands 
($\pm 2$ standard deviations of $\hat{\theta}$, grey error region), where $\tau^2=0.2$ 
and $n=100$. Left to right: variables $1,7,12,16,19$. Top to bottom: $\lambda=0$, $\lambda=0.1$, $\lambda=2$, $\lambda=5$.}
\end{figure}

\begin{figure}[!ht]\centering
\includegraphics[width=\linewidth]{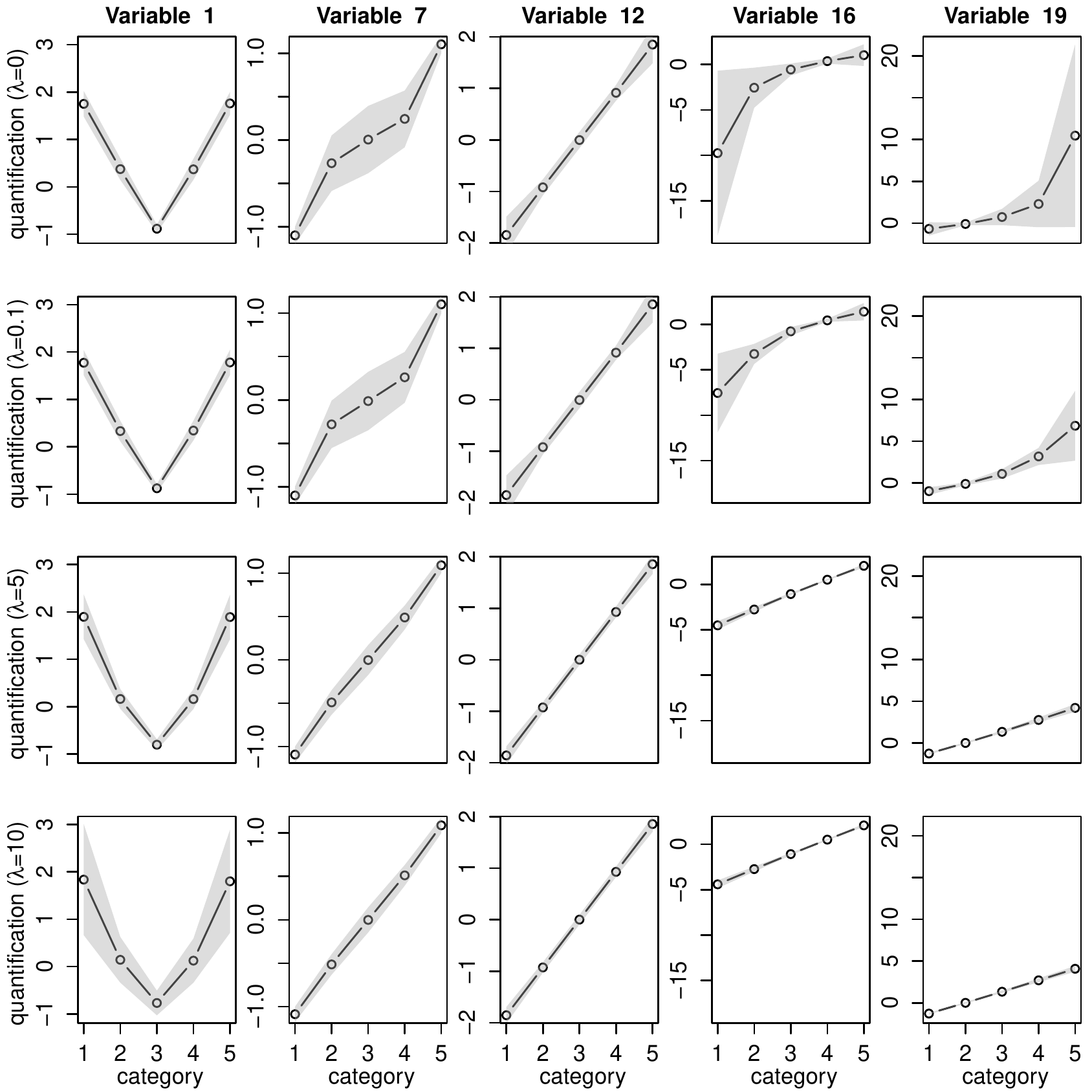}  
\caption{\label{Figure_2} Estimated quantifications (means, black lines) and pointwise uncertainty bands 
($\pm 2$ standard deviations of $\hat{\theta}$, grey error region), where $\tau^2=0.2$ 
and $n=500$. Left to right: variables $1,7,12,16,19$. Top to bottom: $\lambda=0$, $\lambda=0.1$, $\lambda=5$, $\lambda=10$.}
\end{figure}

Since, for simplicity, the same transformation/categorization is applied to all variables within each princial component, we will only present the results for variables $1,7,12,16,19$, which display the V-shaped, S-shaped, linear, square root and quadratic transformation, respectively. Figures~\ref*{Figure_1} and \ref*{Figure_2} show the respective subsets of coefficient vector $\hat{\theta}$ averaged over 500 iterations (black lines) along with pointwise uncertainty bands ($\pm 2$ standard deviations, grey error region), where $\tau^2=0.2$, $n=100$ and $n=500$, respectively. Since both eigenvectors/loadings and quantifications are only defined up to the sign, further restrictions need to be involved when plotting/comparing results across iterations. For (truly) monotone quantifications, we generally require that the overall trend/shape is rather increasing than decreasing, that means, quantifications are multiplied by $-$1 if necessary. For (truly) non-monotone but symmetric(!) quantifications, we require that the shape is V rather than inverted V. Technically speaking, the choice whether to multiply by $-$1 (i.e., flip the fitted quantifications or not) is based on first/second-order differences, i.e., discrete versions of first/second derivative. 


From Figure~\ref*{Figure_1} (column/variable 1) we see that, under true and substantial non-monotonicity, (penalized) non-linear PCA with small $\lambda$ clearly outperforms penalized non-linear PCA with large $\lambda$ regarding both bias and variance. At first glance, this result is counterintuitive having results from smooth, non-linear regression in mind. However, it can be explained as follows. If $\lambda$ is large, scoring rules are forced to be approximately linear with arbitrary orientation. Then, the V-shape transformation (i.e., flip, if necessary) leads to the butterfly-shaped grey/uncertainty regions (Figure~\ref*{Figure_1}, bottom left). But note, suchlike transformation to obtain a (rather) convex scoring rule for variables 1--6 (instead of a transformation yielding (rather) increasing rules for variable 7--20) is necessary to see that small $\lambda$ indeed gives to the true, underlying categorizations. 
Figure~\ref*{Figure_2} with $n=500$ clarifies, that the increase in sample size outweighs the penalty to some extent such that PCA even with $\lambda=10$ is able to detect non-monotonicity, but still suffers from higher variance.  
As before, an explanation is that penalizing deviations from linearity (together with the potential flip to obtain a V-shaped scaling function) still has some influence regardless of the sample size. On the other hand, we see that, under true monotonicity (Figure~\ref*{Figure_1} and \ref*{Figure_2}, columns 2--5), penalized optimal scaling is clearly superior to usual non-linear PCA ($\lambda = 0$) with regard to estimation uncertainty. Even under true non-linearity (variable 7, 16 and 19), purely non-linear PCA seems to be problematic if the number of observations is relatively small. Here, the variance observed reaches values of undesirable amount (Figure~\ref*{Figure_1}, top row). Fortunately, a small change in penalty ($\lambda = 0.1$, second row) is already able to attack the problem, which allows us to capture the non-linear structure while keeping the amount of uncertainty comparatively low. Also if the number of observations is high, penalized non-linear PCA with $\lambda=0.1$ performs best (among the results shown) as it represents a good compromise yielding low variance while keeping the true, possibly non-linear, shape.

\section{Application to Real World Data}\label{Sec4}

While the simulation studies in the previous section provided some illustration on and insight into the proposed method's behavior, the real potential/benefit is only seen from real data applications. For that purpose, we will consider two, publicly available data sets, with the main focus being on the ICF core set for chronic widespread pain already mentioned in Section~\ref*{sectionIntro}.  

\subsection{The ICF Core Set for Chronic Widespread Pain} 
The ICF core set for CWP, available from \texttt{R} package \texttt{ordPens}, contains 420 observations of 67 ordinally scaled variables measuring difficulties in functioning, activities and reduction of quality of life of patients with CWP. Each ICF factor is associated with one of the following four types: `body functions', `body structures', `activities and participation', and `environmental factors'. The latter are measured on a nine-point Likert scale ranging from $-4$ `complete barrier' to $+4$ `complete facilitator'. All remaining factors are evaluated on a five-point Likert scale ranging from $0$ `no problem' to $4$ `complete problem'. Due to space limitations,  we refer to the online supplementary material for an overview on the data analyzed, covering, inter alia, a description of the 67 ICF categories along with observed frequencies (Table~S1, Figures~S1--S2). For more detailed information on the ICF, we refer to \citet{Cieza:2004} and \citet{GerHogObeTut:2011}.

For validating the core set for CWP, ICF evaluations have been compared to the general purpose short-form health survey SF-36~\citep{WarShe:92}. More specifically, this has been done by regressing a physical health component summary calculated from the SF-36 on the ICF data~\citep{GerHogObeTut:2011, Burkner:2020}. The SF-36 summary measures, a physical and a mental health component score, however, were originally constructed by (standard, linear) principal component analysis~\citep{Mchorney:1993}. So for constructing rather disease-specific, ICF-based summary scores, PCA on the ICF data itself appears to be a (more) sensible approach. 

\subsubsection{Smoothing Coefficient Vectors}
Penalized optimal scaling is to be applied to detect (latent) dimensions of symptoms reducing quality of life and related problems of CWP patients as measured by the most important principal components. In analogy to~\cite{Mchorney:1993}, we perform PCA with $m=2$ components, as also resulting from the scree test, see Figure~\ref*{Figure_5} and details as given in Section~\ref*{scree} below. Figure~\ref*{Figure_3} illustrates the estimated coefficients of selected variables for different values of the penalty parameter $\lambda$. The black lines refer to $\lambda \to 0$, the red dashed lines refer to $\lambda=0.5$, and the green dotted lines to $\lambda=5$. It is noticeable that with an increasing penalty parameter quantifications become increasingly linear.  

\begin{figure}[tb]
\begin{center}
\includegraphics[width=\linewidth]{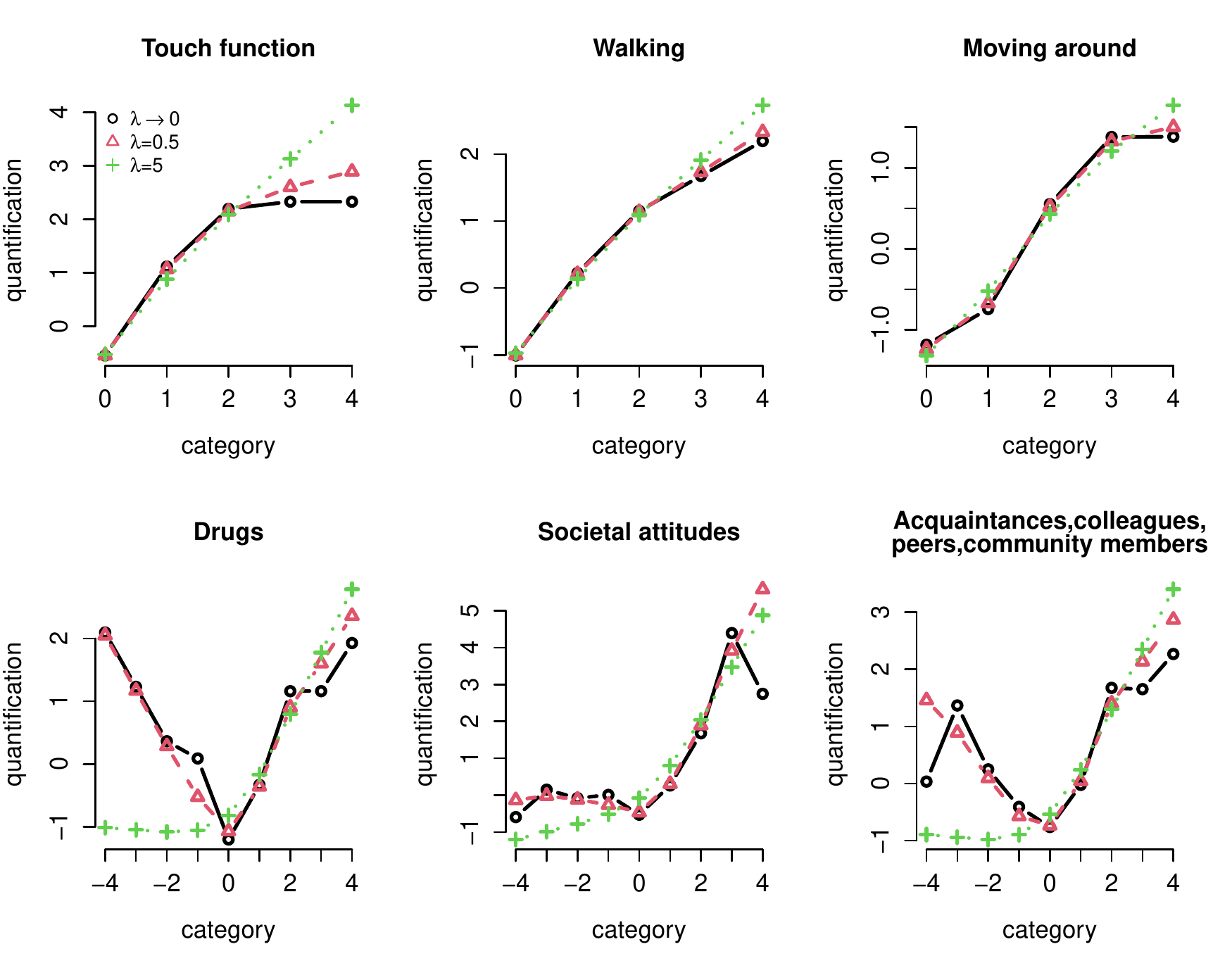} 
\caption{ Category quantifications/scores for $\lambda \to 0$ (solid black), $\lambda=0.5$ (dashed red), 
$\lambda=5$ (dotted green). Monotonicity constraint only applied to variables corresponding to `body functions', `body structures', `activities and participation' (top row). }\label{Figure_3}  
\end{center}
\end{figure} 

For the variable `touch function' in Figure~\ref*{Figure_3}, for instance, the impact of the penalty can be clearly seen with regularization towards linearity; see also variable \emph{b265} (ICF code) in the online appendix (Table~S1, Figure~S3). In general, for all variables of type `body functions', `body structures', or `activities and participation', such as `touch function', `walking' and `moving around' (all of which having coding schemes ranging from `no problem' to `complete problem'), we strict ourselves to monotone scoring rules, as activities (like `moving around') or body functions (like `touch function') are typically affected more and more with worsening CWP. For the scale $-4$ `complete barrier' to $+4$ `complete facilitator' on the other hand, as given with `environmental factors', such restriction would lead in the wrong direction. As observed in Figure~\ref*{Figure_3}~(bottom row), some very non-monotonic effects can be discovered if using a rather small $\lambda$, which can be seen as a clear benefit of non-linear PCA over usual (linear) PCA. When using the penalized method proposed (see, e.g., the red triangles), quantifications are smoother than for unpenalized non-linear PCA (black circles), which is desirable, as wiggly coefficients are hard to interpret.  
Due to a lack of space, results for the other ICF categories are not presented here, but can be found in the online supplements (Figures~S3--S8). 
Finally, it should be noted that $\lambda \to 0$, i.e., $\lambda > 0$ instead of $\lambda = 0$, enables fitting of quantifications (via linear intra/extrapolation) even for levels that are not observed in the data (compare, e.g., level 4 of `walking' in Figure~\ref*{Figure_3}). 

\subsubsection{Smoothing Parameter Selection}\label{icfcv}
\begin{figure}[tb]\centering 
\includegraphics[width=12cm]{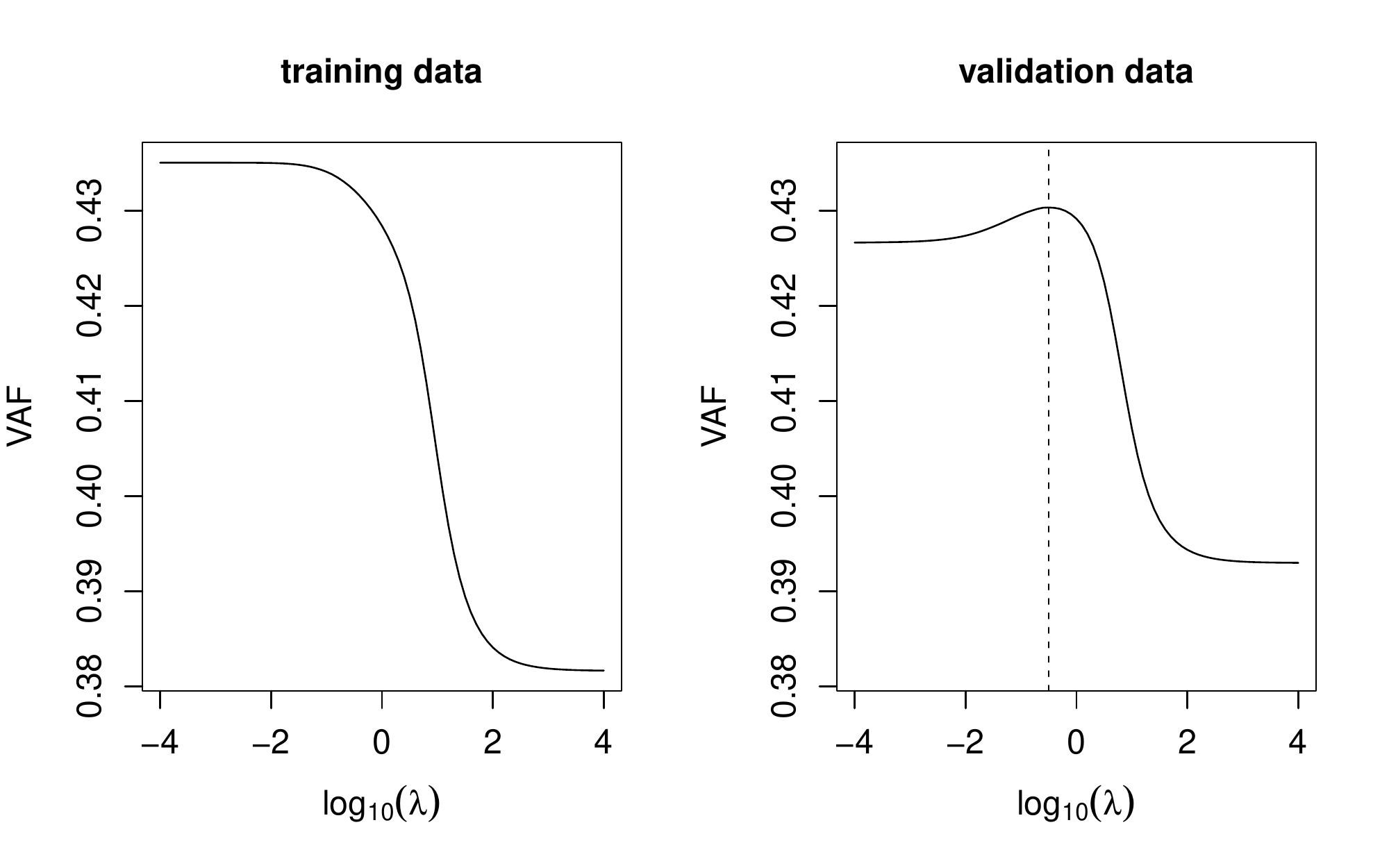}  
\caption{\label{Figure_4} VAF by the first two principal components; left: training sample, right: validation sample.
}
\end{figure}

For evaluating the performance of our approach and selecting the right amount of penalization, we use 5-fold cross-validation as described above. The performance of the quantification rule as given in Figure~\ref*{Figure_4} is measured by the proportion of variance that is explained by the first $m = 2$ principal components when the respective rule is used for scaling the categorical variables.
The optimal smoothing parameter, as determined on the validation set(s), is indicated by the dashed line in Figure~\ref*{Figure_4} (right). Cross-validation results indicate that optimal scaling can indeed be enhanced when using the suggested penalized method. Although penalized scaling functions are less complex, and thus easier to interpret, performance improves on the validation data up to a certain $\lambda$ value and deteriorates from there. Based on those results, we can use $\lambda \approx 0.5$ (where the proportion of variance explained on the validation data reaches its maximum) to find the final scaling rule; compare the dashed vertical line in Figure~\ref*{Figure_4}~(right). Results on training data only, shown in the left panel of Figure~\ref*{Figure_4}, look exactly as sketched in Section~\ref*{lamopt} and indicate virtually the same $\lambda$ value as suggested by cross-validation. That means, an optimal penalty parameter may alternatively be found using the training-data-only approach as described there. Note, the reason why standard, linear PCA (obtained for $\lambda \rightarrow \infty$) appears to perform better on the validation data than the training data is that within 5-fold cross-validation test sets are much smaller than the training set. If sample size reduces while the number of principal components remains constant, however, those components tend to explain a higher proportion of variance than they do on the original data. For illustration, Figure~S9 in the supplementary material shows the proportion of variance accounted for by standard/linear PCA with two components on the ICF data across various random sub-samples of differing size. There it can be seen that average VAF is relatively constant for larger sub-samples but tends to increase at some point if sample size is successively decreased.

\subsubsection{Selecting the Number of Components and Further Interpretation of Results}\label{scree}
\begin{figure}[tb]\centering
	\includegraphics[width=10cm]{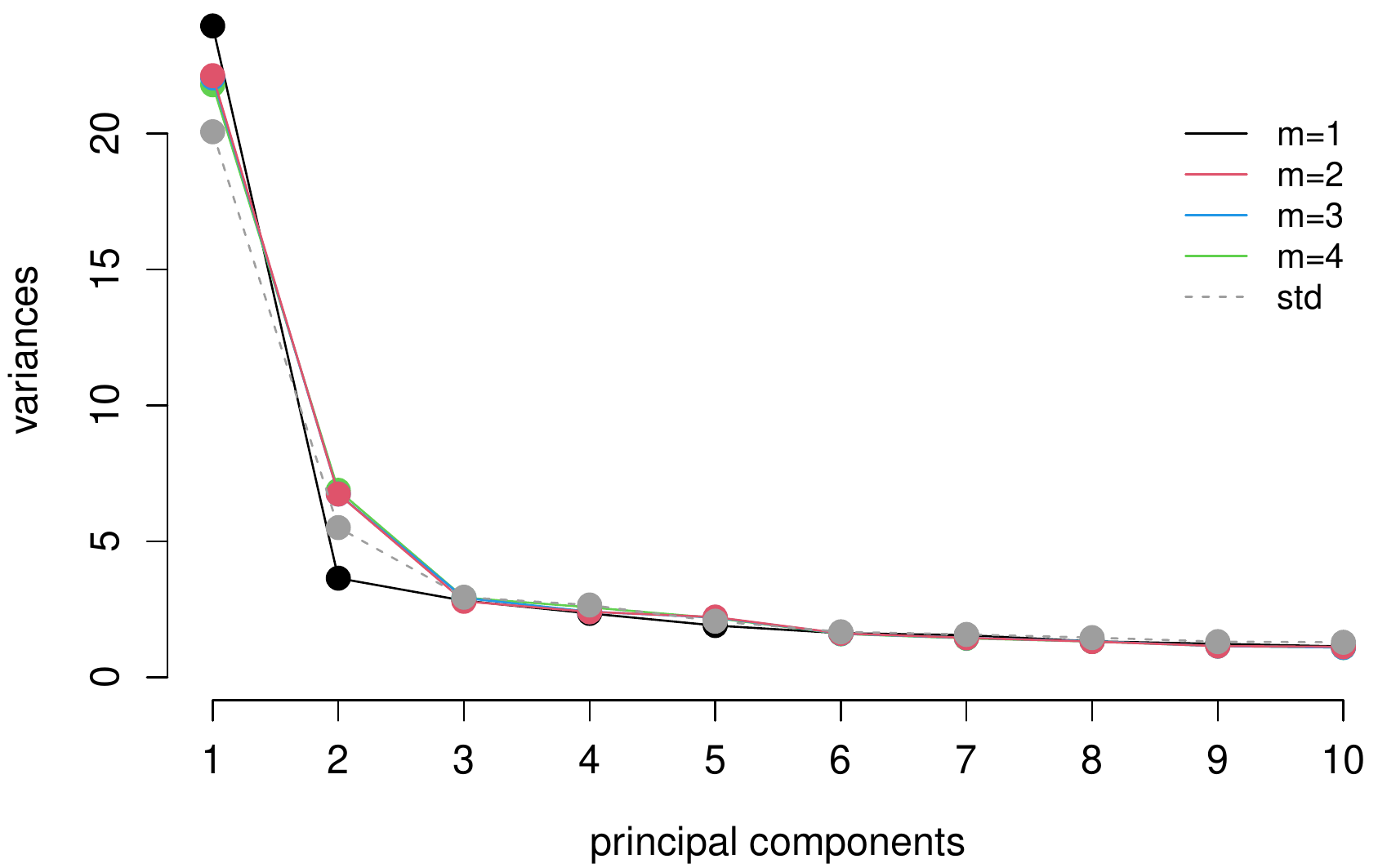} 
	\caption{\label{Figure_5} Scree plots for $m=1$ (solid black), $m=2$ (solid red), $m=3$ (solid blue), $m=4$ (solid green) using the proposed method and standard, linear PCA (dashed grey).}
\end{figure}

To determine a suitable number of principal components, and thus a suitable value for parameter $m$, we consider the approach typically used in the optimal scaling literature (compare Section~\ref{Sec2}), that is, the scree criterion by plotting the PCs against their eigenvalues/variances, and detecting the first break (`elbow'). Figure~\ref*{Figure_5} depicts the scree plot for $m=1$ (black), $m=2$ (red), $m=3$ (blue), and $m=4$ (green) within penalized optimal scaling if using an appropriate $\lambda$ (as chosen via cross-validation, see above) in each case. In addition, we show the scree plot for standard, linear PCA (dashed gray). 
Remember, optimal scaling means standard, linear PCA on the scaled data (compare Section~2). Therefore we can extract also a larger number of components than $m$ (at most $p$), and evaluate the scree plot. For each reasonable $m$, we can expect to see the `elbow' right after the respective position in the scree plot. On the other hand, if the scree plot does not point in the direction of the $m$ value used for estimating the scaling functions, this value appears inadequate compare Section~2 and \citet{LinMeuKooGro:2007}. Indeed, if choosing $m\ge2$, results are quite distinctively in favor of two components, irrespective of the concrete $m\ge2$ being used, which is also the case when employing standard PCA with linear scores. In our experience, as already pointed out (Section~2), it happens quite often that a couple of different $m$ values actually point in the direction of the same number of components to extract according to the scree plot. Additional examples are, for instance, given in the vignette of our R package \texttt{ordPens} (\url{https://htmlpreview.github.io/?https://github.com/ahoshiyar/ordPens/blob/master/vignettes/ordPCA.html}). With the ICF data, however, optimal scaling faces a ``chicken-and-egg problem'', if a decision needs to be made between $m=1$ and $m\ge2$, as the results point towards one component, if choosing $m=1$ for estimating the quantifications. In a case like this, the decision on the number of components (and thus the final $m$ value) should not only be based on the data observed, but also be driven by more substantial considerations (compare, e.g., \citet{Linting:2012}), because eventually a decision has to be made between a one-component solution, i.e., an attempt to compress all information contained in multivariate data into a single number, and allowing for multi-dimensional components. Here, we opted for the latter. Furthermore, we see again (compare Figure~\ref*{Figure_4}, left) that the amount of variance explained by the first two principal components is increased substantially if allowing for non-linear/non-monotone scoring rules, as the first two red circles are well above the gray dots in Figure~\ref{Figure_5}.  
   
The supplementary material gives further insight by providing also the extracted eigenvectors (Table~S2) and the varimax rotated matrix of loadings (Table~S3). In summary, the most important, first principal component (which accounts for roughly 33\% of variance with $\lambda \approx 0.5$ here, compare Figure~\ref*{Figure_5}) can be interpreted as overall CWP condition, with bad health being associated with large values of `body functions', `body structures', or `activities and participation', e.g., `moving around' and `walking'. With respect to `drugs' (\emph{e1101}) both large and small values indicate poor overall condition (compare Figure~\ref*{Figure_3}). In other words, patients with a milder form of CWP are much less affected by the medication, as the latter is much less described as either `barrier' to `facilitator'. It should be pointed out again here, that this would have been impossible to find by standard, linear PCA (or non-linear PCA with monotonicity constraint). Loadings of the other environmental factors are comparatively high (in absolute values) for the second principal component/latent factor. In other words, the second dimension is mainly characterized by those factors. Again, many of those factors show very non-linear or even non-monotone behavior (compare Figure~\ref*{Figure_3} and the online appendix).
More broadly speaking, this example highlights the potential of allowing for non-monotonic effects together with penalization in (ordinal) PCA/optimal scaling. Combining the scree criterion and cross-validation (or some other method for choosing $\lambda$ as described) appears to provide a good compromise between model complexity, fit to the data, and interpretation.

\subsection{Further Example: Depressive Mood Scale Data}\label{ehd}
Next, we will illustrate the proposed method on another publicly available data set for reproducibility. We will investigate the \texttt{ehd} data from \texttt{R} package \texttt{psy} \citep{Falissard:2009}. Example \texttt{R} code for running ordinal PCA as proposed here on the \texttt{ehd} data, is also found in \texttt{ordPens}. 
The \texttt{ehd} data consists of 269 observations of 20 ordinally scaled variables forming a polydimensional rating scale of depressive mood \citep{JouEtal:1988}. Each item is measured on a five-point scale.
\begin{figure}[!ht]
\begin{center}
\includegraphics[width=\linewidth]{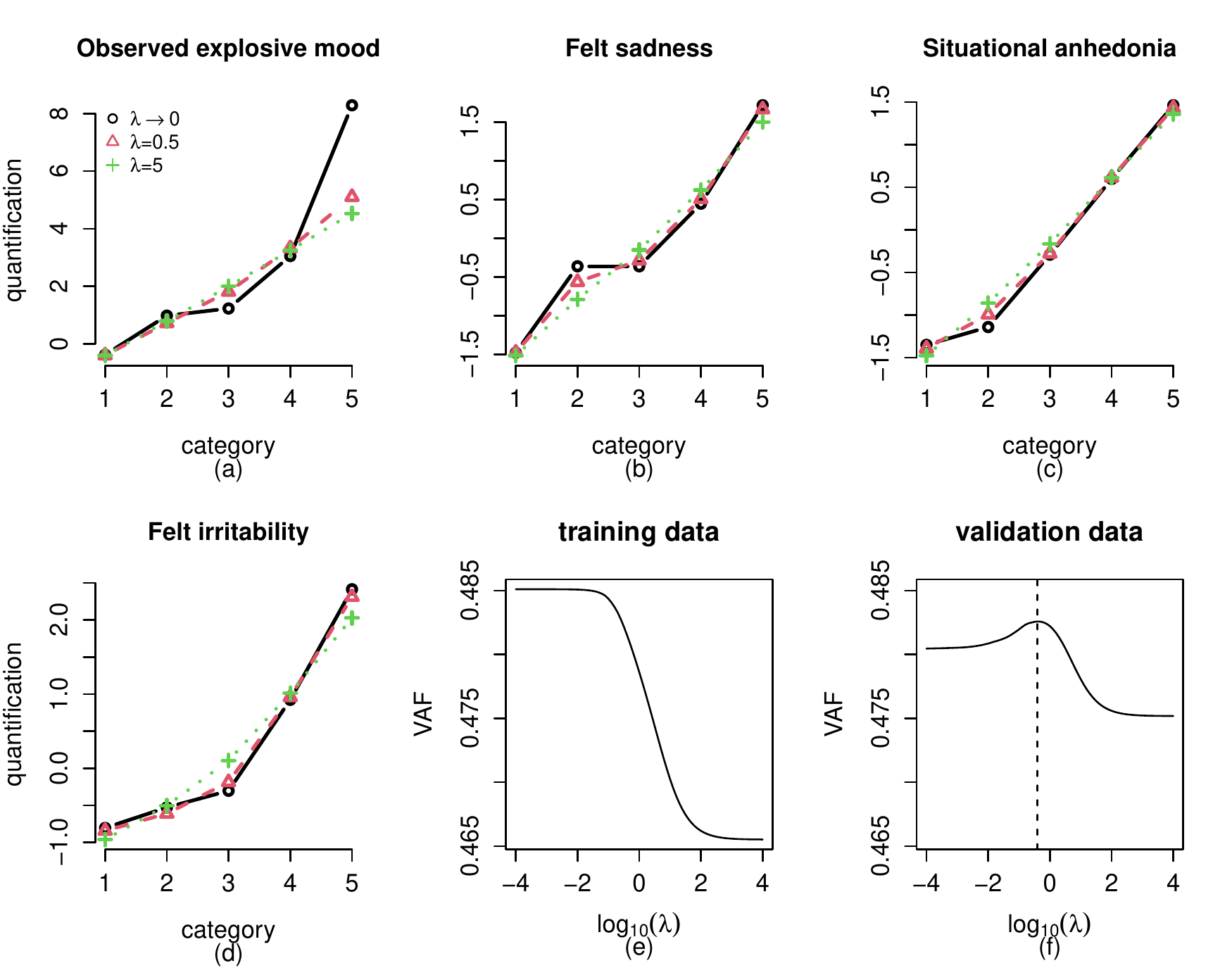} 
\caption{Category quantifications/scores for $\lambda \to 0$ (solid black), $\lambda=0.5$ (dashed red), $\lambda=5$ 
(dotted green) with monotonicity constraint (a)--(d);  
VAF by the first 2 principal components: 
(e) training data, (f) validation data.}\label{Figure_6} 
\end{center}
\end{figure}
We scale the variables by using the presented technique for non-linear PCA with $m=2$ and additional restriction of monotonicity. Figure~\ref*{Figure_6} shows some of the obtained quantifications for different values of smoothing parameter $\lambda$. For the sake of uniformity, the black lines again refer to unpenalized non-linear PCA (i.e., $\lambda \to 0$), the red lines refer to $\lambda=0.5$, and the green lines to $\lambda=5$. As before, it is nicely seen that with larger $\lambda$ quantifications become more and more linear, which is equivalent to standard (linear) PCA using just the category labels.
In this example, a secondary analysis of the data without monotonicity restrictions led to some non-monotone coefficient estimates. As the affected variables of the \texttt{ehd} data are supposed to have consistent, negative association with depressive mood, however, it seems reasonable to assume monotonicity. In other words, content-motivated considerations should be incorporated to enhance interpretability of the estimated effects.
To obtain an optimal amount of smoothing, $\lambda $ was again chosen via 5-fold cross-validation. Figure~\ref*{Figure_6}(e) and (f) shows the mean proportion of variance explained as a function of penalty parameter $\lambda$ (on a logarithmic scale) for both the training data as well as the validation data, respectively. On the training data, this function is of course monotonically decreasing in $\lambda$, as with smaller $\lambda$ more emphasis is put on the data. For $\lambda \to 0$, the original non-linear PCA is obtained, where the explained variance is maximized by construction. On the validation data, however, it's a different story: both (linear) PCA, which is obtained for $\lambda\rightarrow \infty$, and unpenalized/purely non-linear PCA (see $\lambda \rightarrow 0$, i.e., $\log_{10}(\lambda) \rightarrow -\infty$) are worse than penalized, non-linear PCA with $\lambda$-values between $10^{-1} = 0.1$ and $10^{0} = 1$. That means, results of non-linear PCA can be improved by using the penalized fitting algorithm. To obtain a distinct $\lambda$-value for the final scaling rule, cross-validation results as given in Figure~\ref*{Figure_6}(f) can be used as before. Consequently, for the \texttt{ehd} data, we would use $\lambda = 10^{-0.4}\approx 0.5$ (again), where the explained variance on the validation data is maximized (compare the dashed vertical line in Figure~\ref*{Figure_6}(f)).

\section{Summary and Discussion}\label{Sec5}
In the present work, we proposed a new approach to apply principal component analysis on ordinal variables. Those type of data occur often in social and behavioral sciences, but are typically treated as numeric using standard linear PCA, or they are treated as categorical using non-linear PCA (also known as optimal scoring/scaling/quantification). While the former assumes relationships between variables to be linear per construction, the latter can detect non-linear effects but tends to over-fit the data and can lead to estimates that are hard to interpret. To deal with those issues, we presented penalized non-linear PCA as an intermediate between the aforementioned methods where the degree of smoothness can be controlled by a single tuning parameter. We introduced second-order penalties to incorporate the variables' ordinal scale by smoothing out coefficients/quantifications of adjacent levels, more specifically, penalizing non-linearity in the coefficients. The new approach offers both better interpretabiliy of the non-linear transformations of the category labels as well as better performance on validation data than unpenalized optimal scaling. The method proposed is implemented in \texttt{R} and publicly available on CRAN through add-on package \texttt{ordPens}.

Numerical experiments were set up to shed light on the method's capability to parameter recovery and comparison to linear and fully non-linear PCA. We presented a setting involving linear relationships between the variables together with more challenging, non-linear and even non-monotonic effects. Following the concept of bias-variance trade-off, we would associate a large penalty with smaller variance but larger bias (tendency to under-fitting).  
Comparing estimated coefficients, we indeed detected higher variation going along with purely non-linear PCA, especially for small sample size. 
This leads us to the conclusion that it can be preferable to use the proposed penalization technique with an appropriate amount of penalty. Fortunately, already a slight portion of penalization was sufficient to reduce variance while still detecting non-linearity in the scoring rules in our simulations. In practice, cross-validation techniques can be used to find the optimal penalty parameter. The proportion of variance explained on the validation samples also visualizes the potential advantage of penalized optimal scaling against purely non-linear and linear PCA.

To illustrate the application and potential benefits of penalized non-linear PCA in practice, we considered the publicly available ICF core set data for chronic widespread pain.  
We found that two principal components are to be extracted in order to detect the main dimensions of health status in the context of CWP.  
Results on the validation data and non-linearity detected in the majority of quantified variables, signalize the potential impact of non-linear transformations when aiming at dimension reduction, feature extraction and factor selection, compared to both standard, linear and purely non-linear PCA. Penalized optimal scaling as proposed here also allows for monotonicity constraints to enforce monotonic scoring rules/quantifictions. The latter often makes sense as seen with ICF variables of type `body functions', `body structures', and `activities and participation' (or the `depressive mood scale data' also considered). Sometimes, however, non-monotonic transformations provide valuable, additional insight, as it was the case with ICF environmental factors.

\singlespacing 
\bibliography{references}

\newpage
\section*{Supporting Information}
The following supporting information may be found in the online edition of the article: 
\newline
\newline
\fbox{\parbox{\textwidth}{\small{
\textbf{Figure~S1.} Summary for ICF categories corresponding to `body functions', `body structures', `activities and participation' (prefix `b', `d', `s') on individual level.
Coding scheme for categories: 0 `no problem', 1 `mild problem', 2 `moderate problem', 3 `severe problem', 4 `complete problem'.
\newline
\textbf{Figure~S2.} Summary for ICF categories corresponding to `environmental factors' (prefix `e') on individual level.
Coding scheme for categories: -4 `complete barrier',$\ldots$, -1 `mild barrier', 0 `no barrier or facilitator', 1 `mild facilitator',$\ldots$, 4 `complete facilitator'.
\newline
\textbf{Figure~S3.} Category quantifications/scores for $\lambda \to 0$ (solid black), $\lambda=0.5$ (dashed red), $\lambda=5$ (dotted green). Monotonicity constraint only applied to variables corresponding to `body functions', `body structures', `activities and participation' (prefix `b', `d', `s'). 
\newline
\textbf{Figure~S4.} Category quantifications/scores for $\lambda \to 0$ (solid black), $\lambda=0.5$ (dashed red), $\lambda=5$ (dotted green). Monotonicity constraint only applied to variables corresponding to `body functions', `body structures', `activities and participation' (prefix `b', `d', `s'). 
\newline
\textbf{Figure~S5.} Category quantifications/scores for $\lambda \to 0$ (solid black), $\lambda=0.5$ (dashed red), $\lambda=5$ (dotted green). Monotonicity constraint only applied to variables corresponding to `body functions', `body structures', `activities and participation' (prefix `b', `d', `s'). 
\newline
\textbf{Figure~S6.} Category quantifications/scores for $\lambda \to 0$ (solid black), $\lambda=0.5$ (dashed red), $\lambda=5$ (dotted green). Monotonicity constraint only applied to variables corresponding to `body functions', `body structures', `activities and participation' (prefix `b', `d', `s'). 
\newline
\textbf{Figure~S7.} Category quantifications/scores for $\lambda \to 0$ (solid black), $\lambda=0.5$ (dashed red), $\lambda=5$ (dotted green). Monotonicity constraint only applied to variables corresponding to `body functions', `body structures', `activities and participation' (prefix `b', `d', `s'). 
\newline
\textbf{Figure~S8.} Category quantifications/scores for $\lambda \to 0$ (solid black), $\lambda=0.5$ (dashed red), $\lambda=5$ (dotted green). Monotonicity constraint only applied to variables corresponding to `body functions', `body structures', `activities and participation' (prefix `b', `d', `s'). 
\newline
\textbf{Figure~S9.} Proportion of variance accounted for by standard/linear PCA with two components on the ICF data across various random sub-samples of differing size.
\newline
\newline
\textbf{Table~S1.} ICF categories included in the Comprehensive ICF Core Set for chronic widespread pain.
\newline
\textbf{Table~S2.} Eigenvectors of (scaled) ICF data.
\newline
\textbf{Table~S3.} Varimax rotated loadings of ICF data.
The higher loading for each variable is highlighted.
}}}

\end{document}


\title{Penalized Optimal Scaling for Ordinal Variables 
with an Application to International Classification of Functioning Core Sets \\
\vspace{25mm}
Supplementary Material}

  \date{}

\maketitle
\thispagestyle{empty}

\newgeometry{left=35mm, right=35mm, top=30mm, bottom=30mm}

\beginsupplement

\newgeometry{left=35mm, right=35mm, top=20mm, bottom=20mm}

\begin{table}[H]
 \resizebox{\linewidth}{!}{
\centering
\begin{tabular}{llll}
& & & \\ 
  \hline
  & body functions &  & activities and participation \\ 
  \hline
       &  &      &                 \\
 \emph{b122} & Global psychosocial functions &  \emph{d160} & Focusing attention \\ 
 \emph{b126} & Temperament and personality functions &  \emph{d175} & Solving problems \\ 
 \emph{b130} & Energy and drive functions &  \emph{d220} & Undertaking multiple tasks \\ 
 \emph{b134} & Sleep functions &  \emph{d230} & Carrying out daily routine \\ 
 \emph{b140} & Attention functions &  \emph{d240} & Handling stress and other \\ 
 \emph{b147} & Psychomotor functions &  & \quad psychological demands \\ 
 \emph{b152} & Emotional functions &  \emph{d410} & Changing basic body position \\ 
 \emph{b1602} & Content of thought &  \emph{d415} & Maintaining a body position \\ 
 \emph{b164} & Higher level cognitive functions &  \emph{d430} & Lifting and carrying objects \\ 
 \emph{b180} & Experience of self and time functions &\emph{d450} & Walking \\ 
 \emph{b260} & Proprioceptive function &  \emph{d455} & Moving around \\ 
 \emph{b265} & Touch function &  \emph{d470} & Using transportation \\ 
 \emph{b270} & Sensory functions related to &  \emph{d475} & Driving \\ 
       & \quad temperature and other stimuli &  \emph{d510} & Washing oneself \\ 
 \emph{b280} & Sensation of pain &  \emph{d540} & Dressing \\ 
 \emph{b430} & Haematological system functions &  \emph{d570} & Looking after one's health \\ 
 \emph{b455} & Exercise tolerance functions &  \emph{d620} & Acquisition of goods and services \\ 
 \emph{b640} & Sexual functions &  \emph{d640} & Doing housework \\ 
 \emph{b710} & Mobility of joint functions &  \emph{d650} & Caring for household objects \\ 
 \emph{b730} & Muscle power functions &  \emph{d660} & Assisting others \\ 
 \emph{b735} & Muscle tone functions &  \emph{d720} & Complex interpersonal interactions \\ 
 \emph{b740} & Muscle endurance functions &  \emph{d760} & Family relationships \\ 
 \emph{b760} & Control of voluntary movement functions &  \emph{d770} & Intimate relationships \\ 
 \emph{b780} & Sensations related to muscles and  &  \emph{d845} & Acquiring, keeping and  \\ 
       & \quad movement functions        &   & \quad terminating a job \\
       &  &  \emph{d850} & Remunerative employment \\ 
       &  &  \emph{d855} & Non-remunerative employment \\ 
       &  &  \emph{d910} & Community life \\ 
       &  &  \emph{d920} & Recreation and leisure \\ 
       &  &      &                 \\
     \hline
       & environmental factors & & body structures  \\ 
     \hline
       &  &      &                 \\
 \emph{e1101} & Drugs & \emph{s770} & Additional musculoskeletal  \\ 
\emph{e310} & Immediate family &  & \quad structures related to movement \\ 
\emph{e325} & Acquaintances, peers, colleagues,  &  &  \\ 
       & \quad neighbours and community members &  &  \\ 
\emph{e355} & Health professionals &  &  \\ 
\emph{e410} & Individual attitudes of immediate family  &  &  \\ 
       & \quad members              &  &  \\
\emph{e420} & Individual attitudes of friends &  &  \\ 
\emph{e425} & Individual attitudes acquaintances, peers,  &  &  \\ 
       & \quad colleagues, neighbours and community  &  &  \\ 
       & \quad members              &  &  \\
\emph{e430} & Individual attitudes of people in positions  &  &  \\ 
       & \quad of authority         &  &   \\
\emph{e450} & Individual attitudes of health professionals &  &  \\ 
\emph{e455} & Individual attitudes of health-related  &  &  \\ 
       & \quad professionals      &  &  \\
\emph{e460} & Societal attitudes &  &  \\ 
\emph{e465} & Social norms, practices and ideologies &  &  \\ 
\emph{e570} & Social security services, systems and  &  &  \\ 
       & \quad policies           &  &  \\
\emph{e575} & General social support services, systems  &  &  \\  
       & \quad and policies       &  &  \\
\emph{e580} & Health services, systems and policies &  &  \\ 
\emph{e590} & Labour and employment services,  &  &  \\ 
       & systems and policies    &   &  \\
   \hline
\end{tabular}
}
\caption{\label{Table_S1} ICF categories included in the Comprehensive ICF Core Set for chronic widespread pain.
}
\end{table}

 
\begin{figure}[H]\centering 
\includegraphics[width=0.825\linewidth]{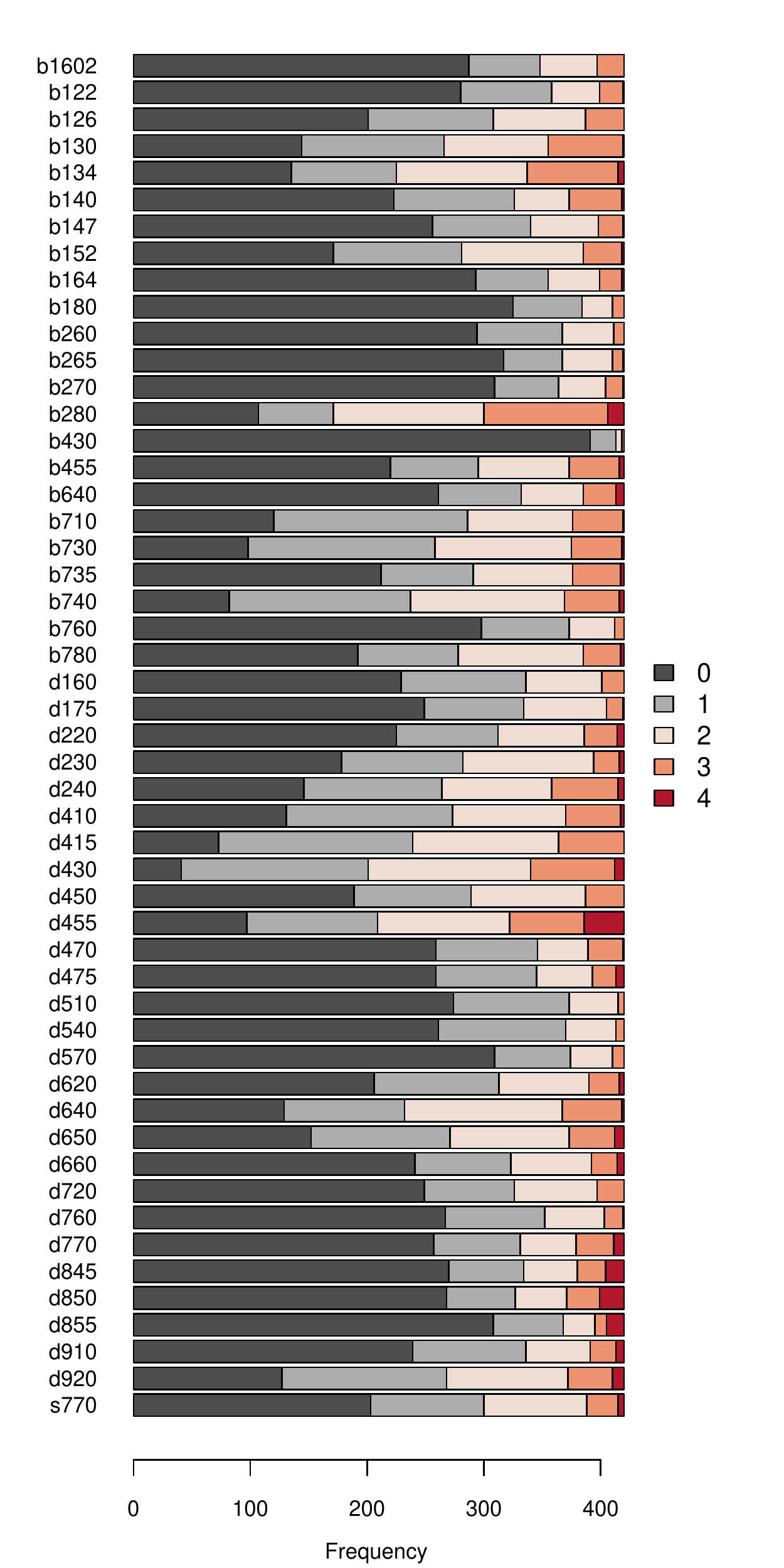} 
\caption{\label{Figure_S1} Summary for ICF categories corresponding to `body functions', `body structures', `activities and participation' (prefix `b', `d', `s') on individual level.
Coding scheme for categories: 0 `no problem', 1 `mild problem', 2 `moderate problem', 3 `severe problem', 4 `complete problem'.
}
\end{figure}

\begin{figure}[H]\centering
\includegraphics[width=0.825\linewidth]{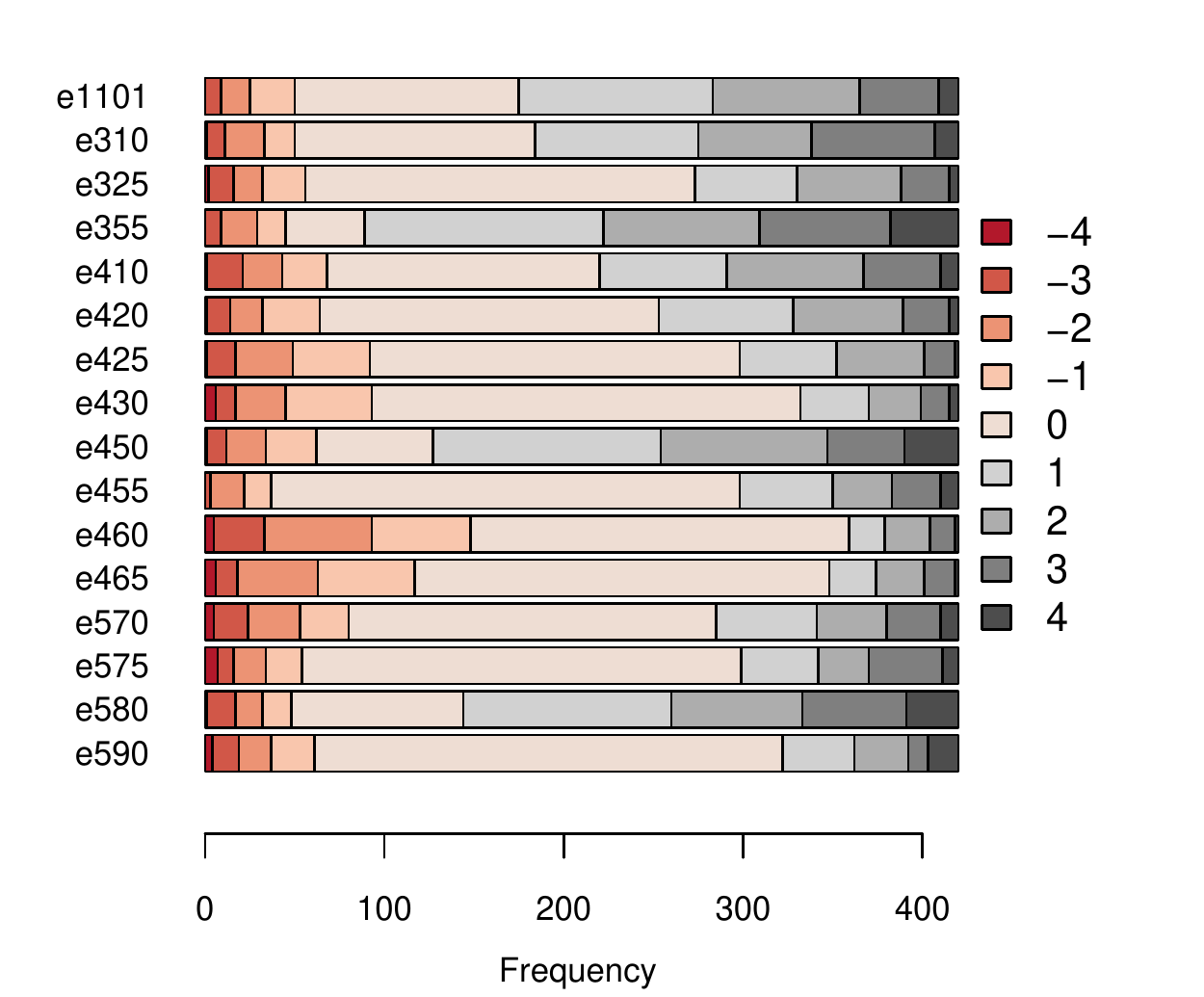} 
\caption{\label{Figure_S2} Summary for ICF categories corresponding to `environmental factors' (prefix `e') on individual level.
Coding scheme for categories: -4 `complete barrier',..., -1 `mild barrier', 0 `no barrier or facilitator', 1 `mild facilitator',..., 4 `complete facilitator'.
}
\end{figure}

                  \newgeometry{left=35mm, right=35mm, top=30mm, bottom=30mm}

\begin{figure}[H] 
\centering
\includegraphics[width=\linewidth]{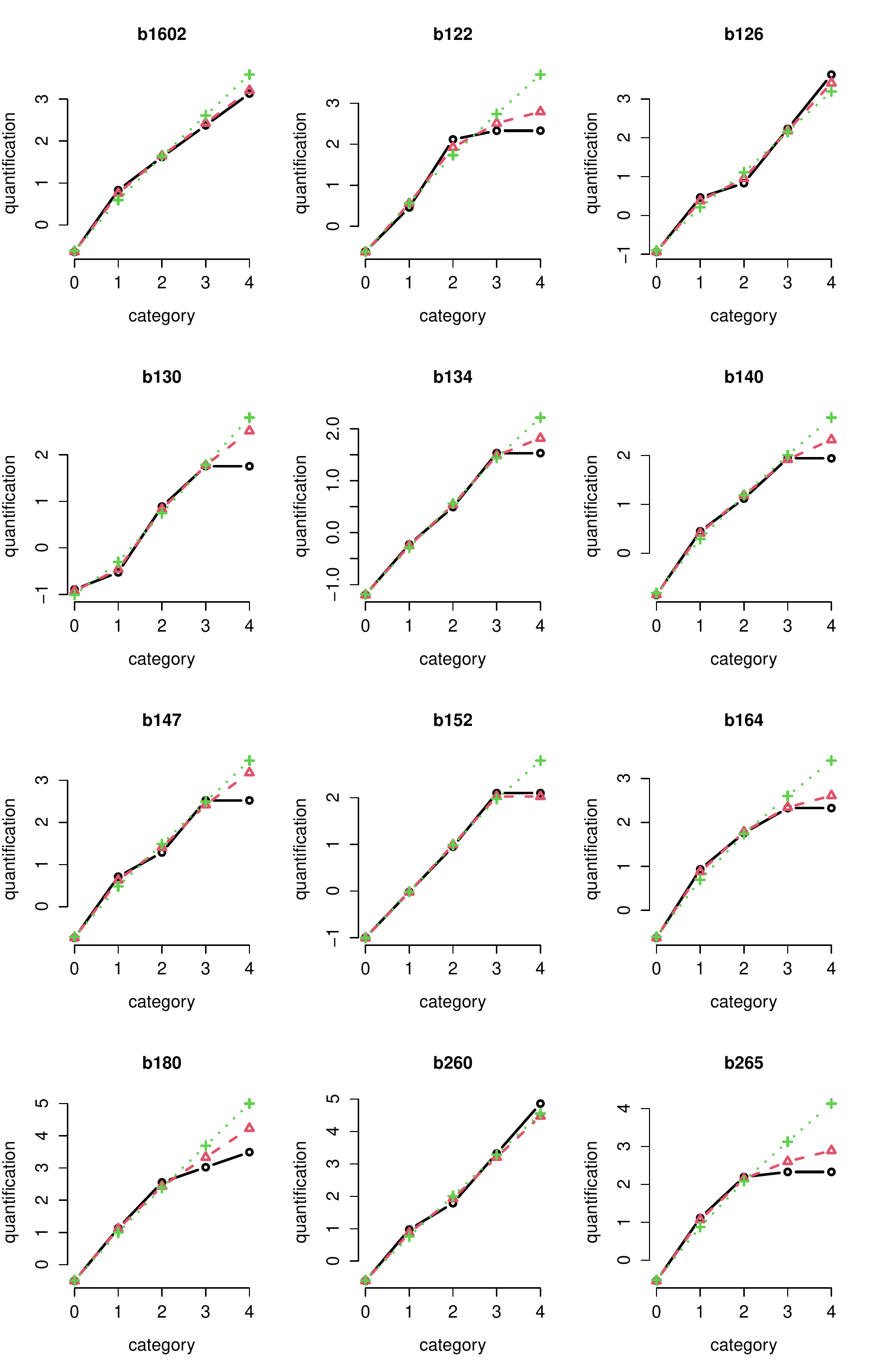}
\caption{\label{Figure_S3} Category quantifications/scores for $\lambda \to 0$ (solid black), $\lambda=0.5$ (dashed red), 
$\lambda=5$ (dotted green). Monotonicity constraint only applied to variables corresponding to `body functions', `body structures', `activities and participation' (prefix `b', `d', `s'). 
}

\end{figure}

\begin{figure}[H]
\centering
\includegraphics[width=\linewidth]{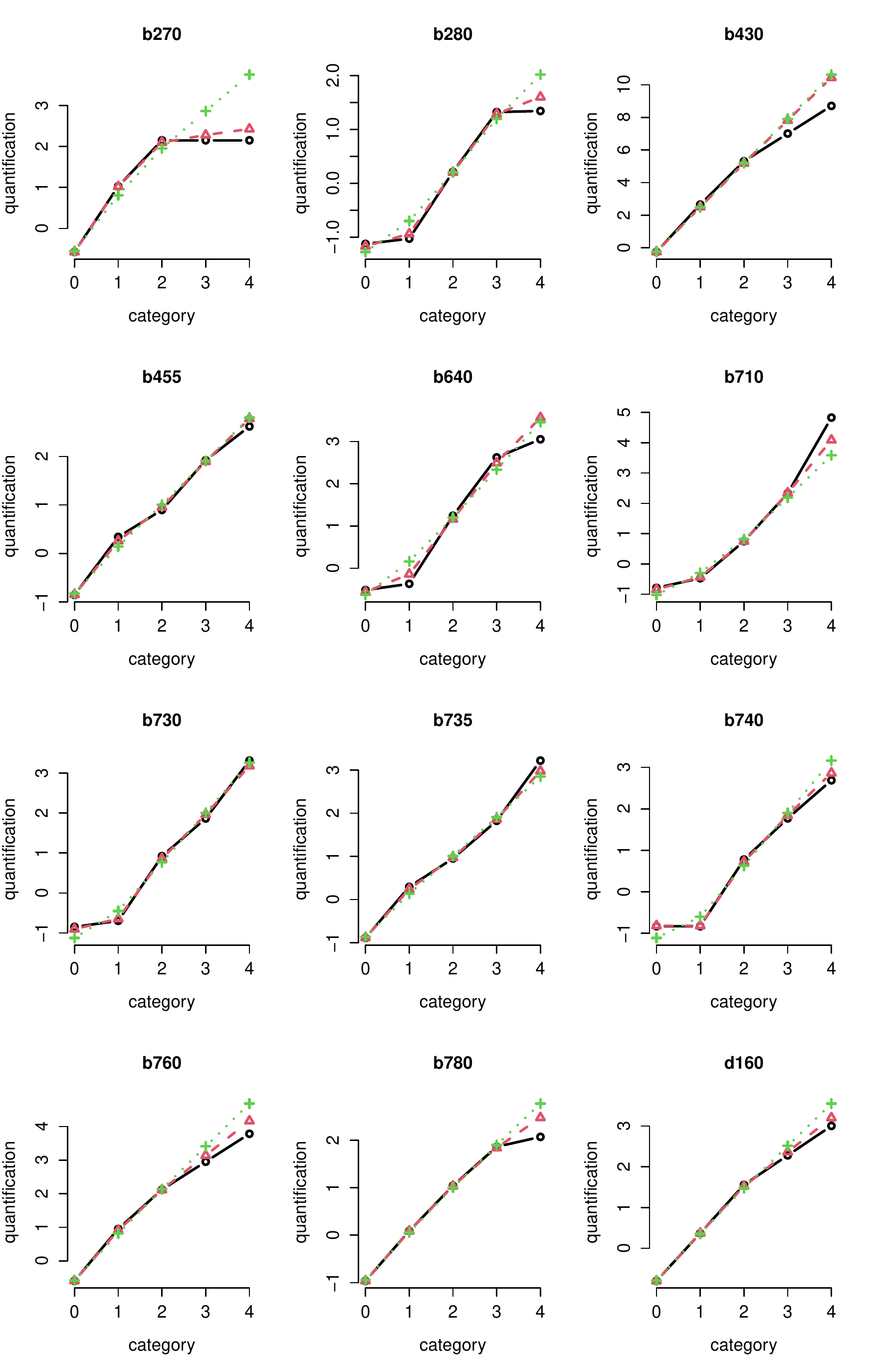}
\caption{\label{Figure_S4} Category quantifications/scores for $\lambda \to 0$ (solid black), $\lambda=0.5$ (dashed red), 
$\lambda=5$ (dotted green). Monotonicity constraint only applied to variables corresponding to `body functions', `body structures', `activities and participation' (prefix `b', `d', `s'). 
}
\end{figure}

\begin{figure}[H]
\centering
\includegraphics[width=\linewidth]{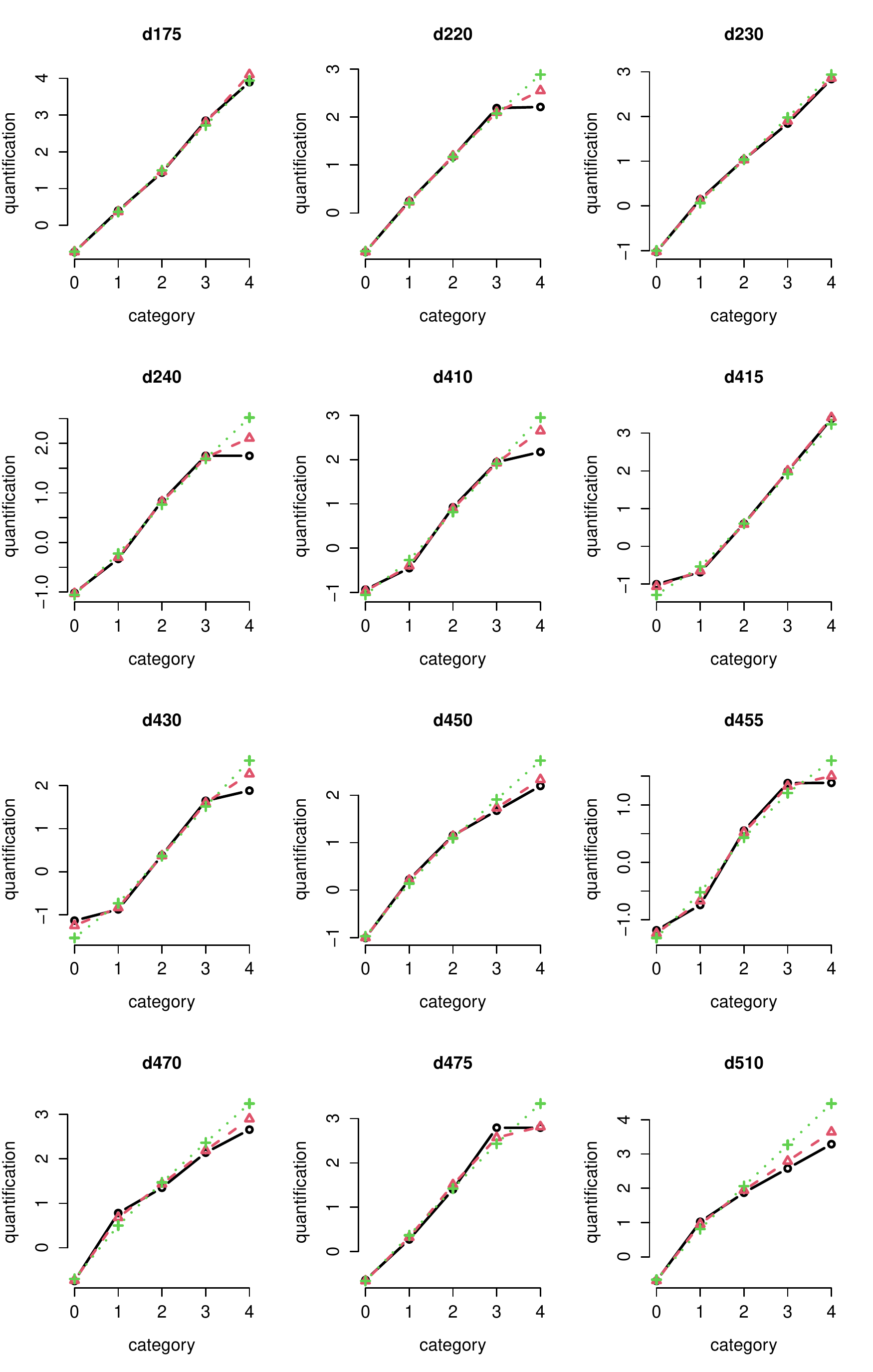}
\caption{\label{Figure_S5} Category quantifications/scores for $\lambda \to 0$ (solid black), $\lambda=0.5$ (dashed red), 
$\lambda=5$ (dotted green). Monotonicity constraint only applied to variables corresponding to `body functions', `body structures', `activities and participation' (prefix `b', `d', `s'). 
}
\end{figure}

\begin{figure}[H]
\centering
\includegraphics[width=\linewidth]{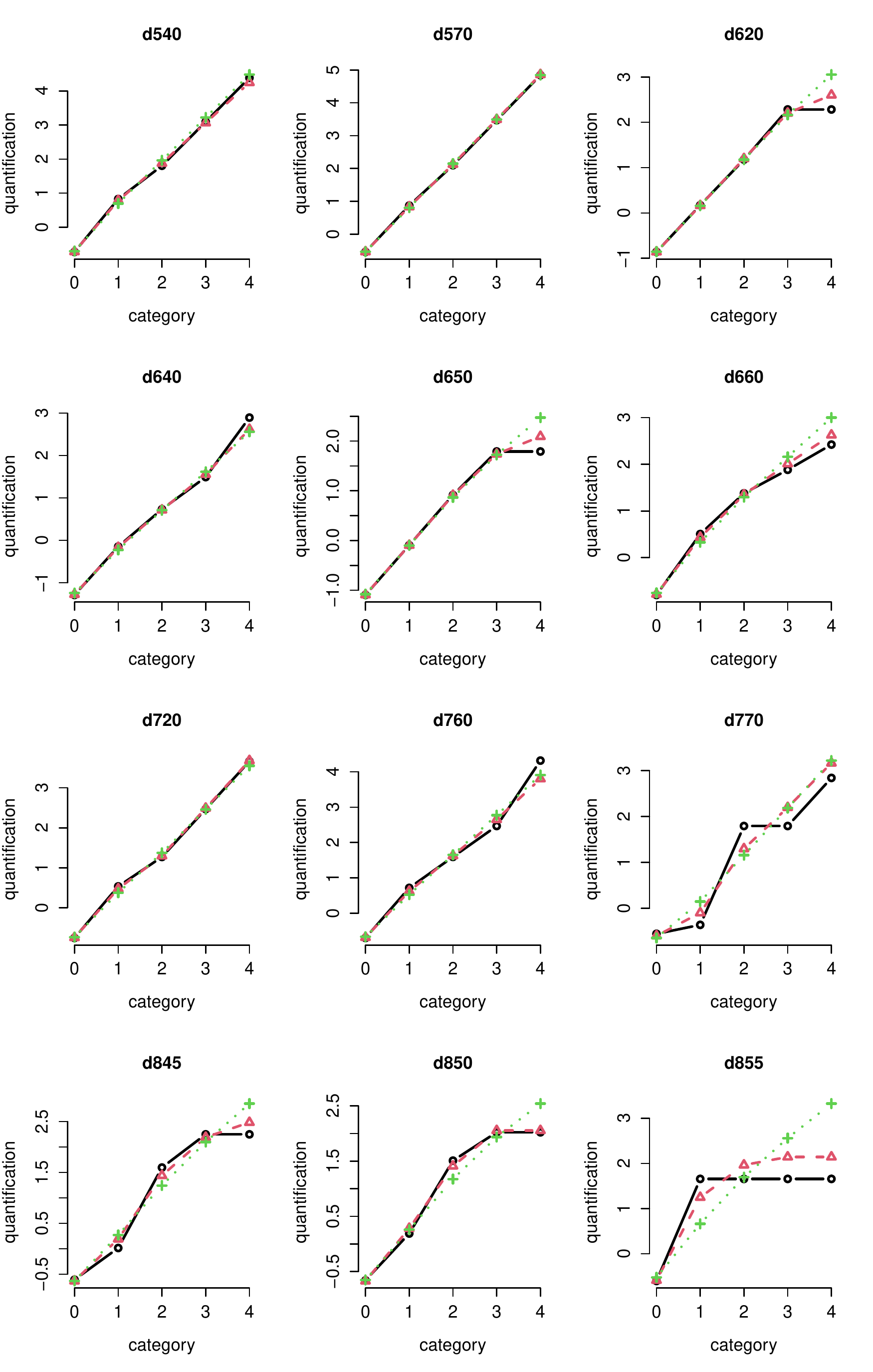}
\caption{\label{Figure_S6} Category quantifications/scores for $\lambda \to 0$ (solid black), $\lambda=0.5$ (dashed red), 
$\lambda=5$ (dotted green). Monotonicity constraint only applied to variables corresponding to `body functions', `body structures', `activities and participation' (prefix `b', `d', `s'). 
}
\end{figure}

\begin{figure}[H]
\centering
\includegraphics[width=\linewidth]{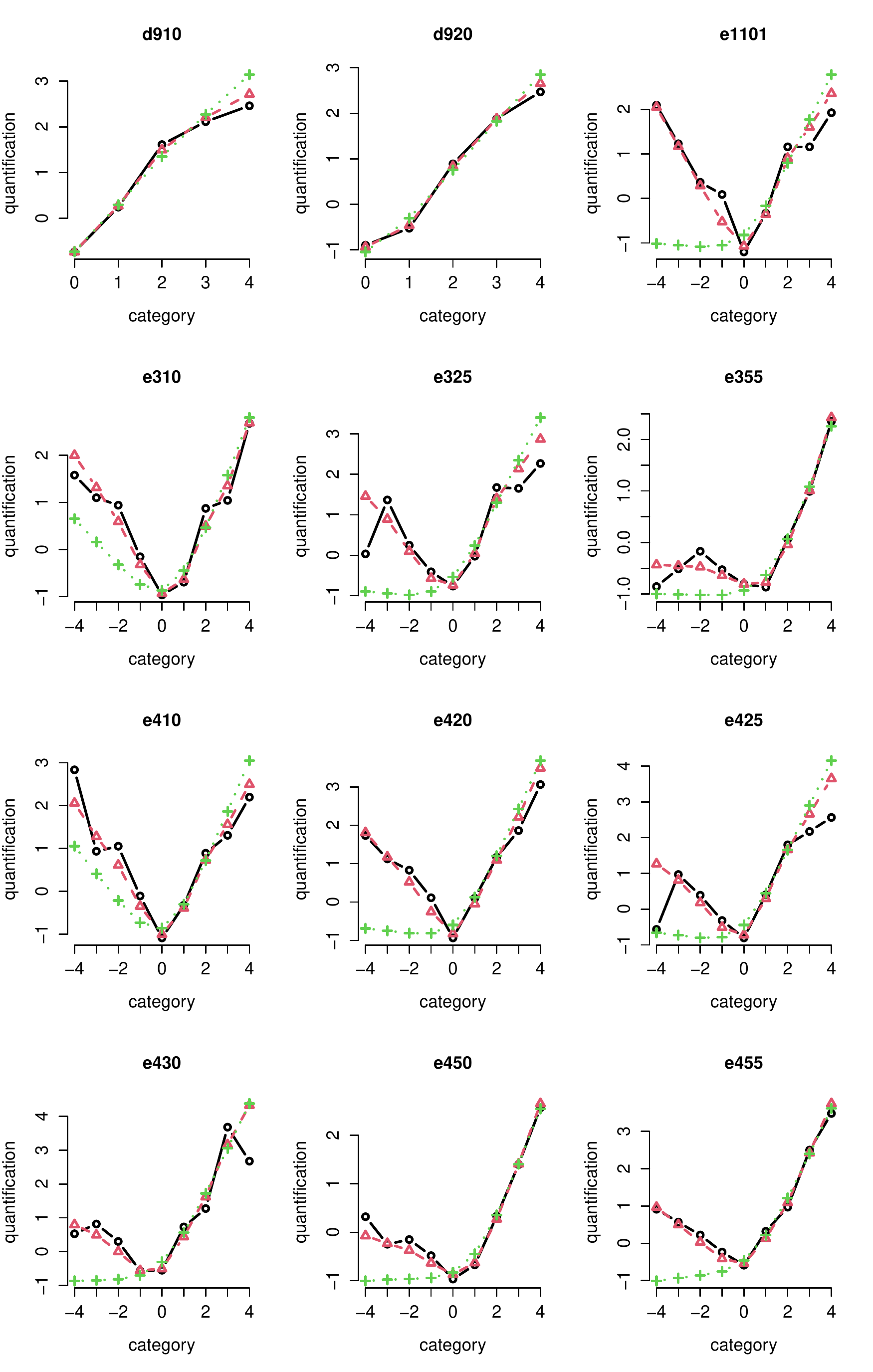}
\caption{\label{Figure_S7} Category quantifications/scores for $\lambda \to 0$ (solid black), $\lambda=0.5$ (dashed red), 
$\lambda=5$ (dotted green). Monotonicity constraint only applied to variables corresponding to `body functions', `body structures', `activities and participation' (prefix `b', `d', `s'). 
}
\end{figure}

\begin{figure}[H]
\centering
\includegraphics[width=\linewidth]{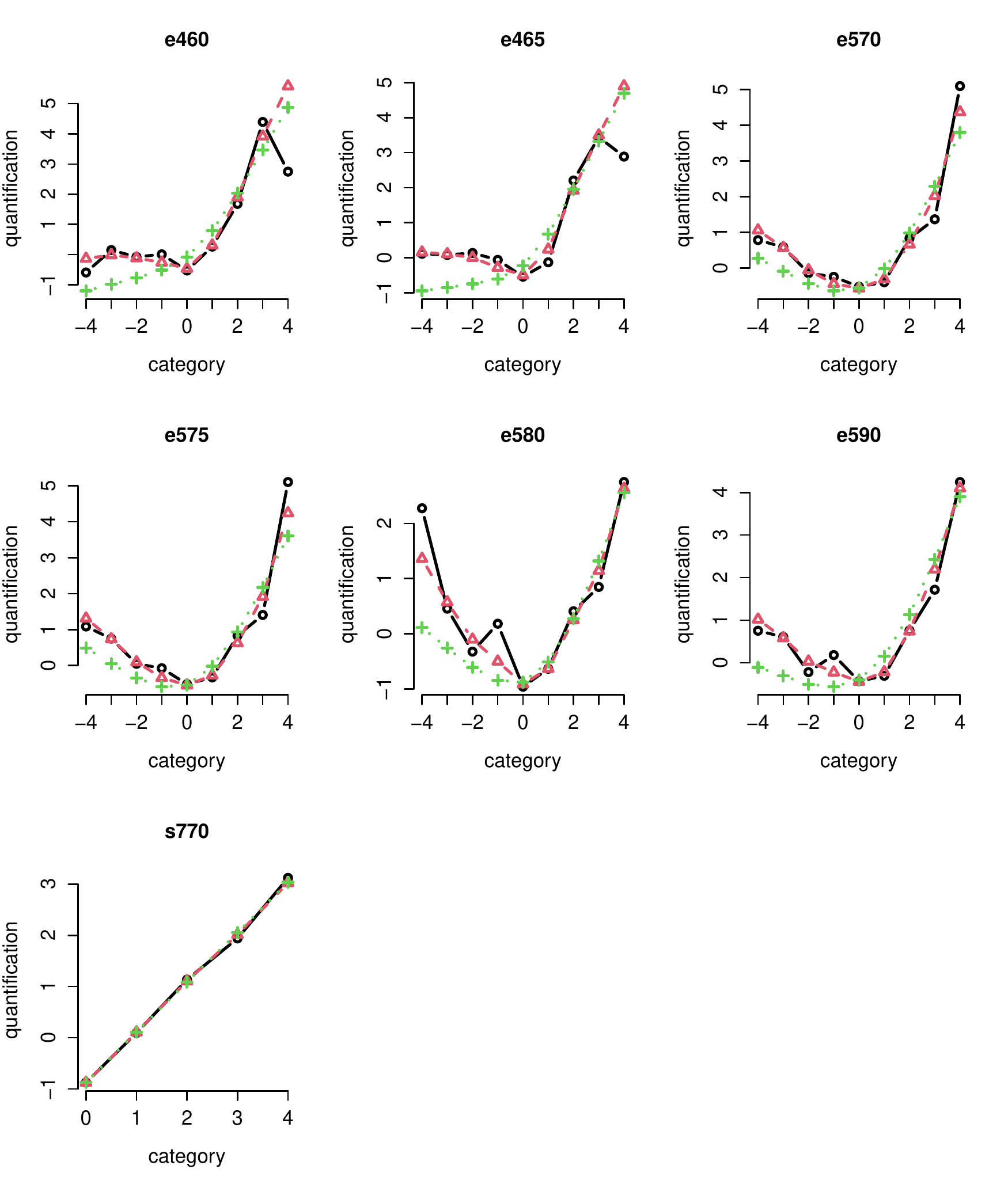}
\caption{\label{Figure_S8} Category quantifications/scores for $\lambda \to 0$ (solid black), $\lambda=0.5$ (dashed red), 
$\lambda=5$ (dotted green). Monotonicity constraint only applied to variables corresponding to `body functions', `body structures', `activities and participation' (prefix `b', `d', `s'). 
}
\end{figure}

\begin{figure}[H]
\centering
\includegraphics[width=\linewidth]{Figure_S9.pdf}
\caption{\label{Figure_S9} Proportion of variance accounted for by standard/linear PCA with two components on the ICF data across various random sub-samples of differing size. It can be seen that average VAF (red line) is relatively constant for larger sub-samples but tends to increase at some point if sample size is successively decreased. The boxplot that roughly corresponds to the size of the validation sets within 5-fold cross-validation on the ICF data (84) is marked in blue. The size of the corresponding training sets within 5-fold cross-validation was 336. From a more general perspective, the finding that VAF tends to increase (at some point) for decreasing (sub-)sample size can be explained/motivated as follows. In the extreme case with just three observations, for instance, two principal components would generally be enough to explain 100\% of the variance (also if $p > 2$), because two-component PCA means `projecting' into a two-dimensional subspace which is spanned by those three observations. With four observations, two components would generally explain more than 66\% of the variance (typically much more, but less then 100\%), and so on.
}
\end{figure}

\begin{table}[H]
\centering
\begin{tabular}{lrr|lrr}
 & PC1 & PC2 &       & PC1 & PC2 \\ 
  \hline  
\emph{b1602} & 0.13 & 0.09 & \emph{d475} & 0.10 & 0.04 \\ 
\emph{b122} & 0.12 & 0.02 & \emph{d510} & 0.13 & 0.04 \\ 
\emph{b126} & 0.14 & 0.05 & \emph{d540} & 0.13 & 0.06 \\ 
\emph{b130} & 0.15 & 0.09 & \emph{d570} & 0.11 & -0.01 \\ 
\emph{b134} & 0.15 & -0.01 & \emph{d620} & 0.16 & 0.02 \\ 
\emph{b140} & 0.15 & 0.10 & \emph{d640} & 0.16 & -0.02 \\ 
\emph{b147} & 0.14 & 0.04 & \emph{d650} & 0.15 & 0.00 \\ 
\emph{b152} & 0.16 & 0.03 & \emph{d660} & 0.12 & -0.02 \\ 
\emph{b164} & 0.14 & 0.07 & \emph{d720} & 0.16 & -0.04 \\ 
\emph{b180} & 0.11 & 0.07 & \emph{d760} & 0.13 & -0.04 \\ 
\emph{b260} & 0.12 & 0.04 & \emph{d770} & 0.10 & -0.05 \\ 
\emph{b265} & 0.13 & 0.04 & \emph{d845} & 0.11 & 0.03 \\ 
\emph{b270} & 0.12 & 0.03 & \emph{d850} & 0.11 & 0.01 \\ 
\emph{b280} & 0.15 & 0.09 & \emph{d855} & 0.07 & -0.10 \\ 
\emph{b430} & 0.04 & 0.01 & \emph{d910} & 0.12 & 0.02 \\ 
\emph{b455} & 0.13 & -0.04 & \emph{d920} & 0.13 & 0.04 \\ 
\emph{b640} & 0.10 & 0.04 & \emph{e1101} & 0.10 & -0.08 \\ 
\emph{b710} & 0.12 & 0.11 & \emph{e310} & 0.11 & -0.14 \\ 
\emph{b730} & 0.15 & 0.10 & \emph{e325} & 0.09 & -0.18 \\ 
\emph{b735} & 0.15 & 0.07 & \emph{e355} & 0.07 & -0.25 \\ 
\emph{b740} & 0.14 & 0.10 & \emph{e410} & 0.12 & -0.13 \\ 
\emph{b760} & 0.11 & 0.08 & \emph{e420} & 0.11 & -0.15 \\ 
\emph{b780} & 0.14 & 0.04 & \emph{e425} & 0.09 & -0.21 \\ 
\emph{d160} & 0.15 & 0.06 & \emph{e430} & 0.06 & -0.23 \\ 
\emph{d175} & 0.15 & 0.01 & \emph{e450} & 0.07 & -0.26 \\ 
\emph{d220} & 0.16 & 0.01 & \emph{e455} & 0.07 & -0.26 \\ 
\emph{d230} & 0.16 & -0.02 & \emph{e460} & 0.05 & -0.26 \\ 
\emph{d240} & 0.16 & 0.03 & \emph{e465} & 0.05 & -0.25 \\ 
\emph{d410} & 0.14 & 0.09 & \emph{e570} & 0.04 & -0.29 \\ 
\emph{d415} & 0.13 & 0.10 & \emph{e575} & 0.04 & -0.30 \\ 
\emph{d430} & 0.14 & 0.04 & \emph{e580} & 0.07 & -0.25 \\ 
\emph{d450} & 0.14 & 0.06 & \emph{e590} & 0.04 & -0.28 \\ 
\emph{d455} & 0.14 & 0.02 & \emph{s770} & 0.12 & 0.06 \\ 
\emph{d470} & 0.13 & -0.04 &  &   &   \\ 
\end{tabular}
\caption{\label{Table_S2} Eigenvectors of (scaled) ICF data.}
\end{table}


\begin{table}[H]
\centering
\begin{tabular}{lrr|lrr}
 & PC1 & PC2 &       & PC1 & PC2 \\ 
  \hline  
\emph{b1602} & $\mathbf{0.65}$ & 0.01 & \emph{d475} & $\mathbf{0.46}$ & -0.06 \\ 
\emph{b122} & $\mathbf{0.56}$ & -0.15 & \emph{d510} & $\mathbf{0.60}$ & -0.11 \\ 
\emph{b126} & $\mathbf{0.68}$ & -0.10 & \emph{d540} & $\mathbf{0.65}$ & -0.07 \\ 
\emph{b130} & $\mathbf{0.75}$ & -0.03 & \emph{d570} & $\mathbf{0.49}$ & -0.22 \\ 
\emph{b134} & $\mathbf{0.66}$ & -0.27 & \emph{d620} & $\mathbf{0.72}$ & -0.22 \\ 
\emph{b140} & $\mathbf{0.75}$ & 0.01 & \emph{d640} & $\mathbf{0.69}$ & -0.31 \\ 
\emph{b147} & $\mathbf{0.64}$ & -0.11 & \emph{d650} & $\mathbf{0.65}$ & -0.23 \\ 
\emph{b152} & $\mathbf{0.73}$ & -0.19 & \emph{d660} & $\mathbf{0.50}$ & -0.24 \\ 
\emph{b164} & $\mathbf{0.70}$ & -0.06 & \emph{d720} & $\mathbf{0.66}$ & -0.35 \\ 
\emph{b180} & $\mathbf{0.57}$ & -0.00 & \emph{d760} & $\mathbf{0.55}$ & -0.30 \\ 
\emph{b260} & $\mathbf{0.55}$ & -0.10 & \emph{d770} & $\mathbf{0.41}$ & -0.29 \\ 
\emph{b265} & $\mathbf{0.62}$ & -0.12 & \emph{d845} & $\mathbf{0.52}$ & -0.11 \\ 
\emph{b270} & $\mathbf{0.57}$ & -0.11 & \emph{d850} & $\mathbf{0.48}$ & -0.14 \\ 
\emph{b280} & $\mathbf{0.73}$ & -0.01 & \emph{d855} & 0.25 & $\mathbf{-0.35}$ \\ 
\emph{b430} & $\mathbf{0.21}$ & -0.04 & \emph{d910} & $\mathbf{0.55}$ & -0.15 \\ 
\emph{b455} & $\mathbf{0.53}$ & -0.30 & \emph{d920} & $\mathbf{0.63}$ & -0.11 \\ 
\emph{b640} & $\mathbf{0.46}$ & -0.05 & \emph{e1101} & $\mathbf{0.38}$ & -0.35 \\ 
\emph{b710} & $\mathbf{0.61}$ & 0.08 & \emph{e310} & 0.36 & $\mathbf{-0.51}$ \\ 
\emph{b730} & $\mathbf{0.73}$ & -0.00 & \emph{e325} & 0.23 & $\mathbf{-0.59}$ \\ 
\emph{b735} & $\mathbf{0.72}$ & -0.07 & \emph{e355} & 0.09 & $\mathbf{-0.73}$ \\ 
\emph{b740} & $\mathbf{0.73}$ & 0.02 & \emph{e410} & 0.39 & $\mathbf{-0.51}$ \\ 
\emph{b760} & $\mathbf{0.57}$ & 0.01 & \emph{e420} & 0.33 & $\mathbf{-0.54}$ \\ 
\emph{b780} & $\mathbf{0.64}$ & -0.11 & \emph{e425} & 0.23 &$\mathbf{ -0.65}$ \\ 
\emph{d160} & $\mathbf{0.73}$ & -0.09 & \emph{e430} & 0.07 & $\mathbf{-0.66}$ \\ 
\emph{d175} & $\mathbf{0.65}$ & -0.22 & \emph{e450} & 0.09 & $\mathbf{-0.76}$ \\ 
\emph{d220} & $\mathbf{0.72}$ & -0.23 & \emph{e455} & 0.06 & $\mathbf{-0.75}$ \\ 
\emph{d230} & $\mathbf{0.69}$ & -0.30 & \emph{e460} & -0.00 & $\mathbf{-0.70}$ \\ 
\emph{d240} & $\mathbf{0.74}$ & -0.17 & \emph{e465} & -0.01 & $\mathbf{-0.69}$ \\ 
\emph{d410} & $\mathbf{0.70}$ & -0.01 & \emph{e570} & -0.10 & $\mathbf{-0.76}$ \\ 
\emph{d415} & $\mathbf{0.68}$ & 0.03 & \emph{e575} & -0.08 & $\mathbf{-0.79}$ \\ 
\emph{d430} & $\mathbf{0.65}$ & -0.11 & \emph{e580} & 0.08 & $\mathbf{-0.72}$ \\ 
\emph{d450} & $\mathbf{0.67}$ & -0.09 & \emph{e590} & -0.09 & $\mathbf{-0.74}$ \\ 
\emph{d455} & $\mathbf{0.62}$ & -0.18 & \emph{s770} & $\mathbf{0.60}$ & -0.04 \\ 
\emph{d470} & $\mathbf{0.54}$ & -0.31 &  &   &   \\ 
\end{tabular} 
\caption{\label{Table_S3} Varimax rotated loadings of ICF data.
The higher loading for each variable is highlighted.}
\end{table}